\documentclass[sigconf]{acmart}
\usepackage{multirow}
\usepackage{float}
\usepackage{url}
\usepackage{algorithm}
\usepackage{algpseudocode}
\usepackage{bm}
\usepackage{graphicx}
\usepackage{caption}
\usepackage[skip=0cm,list=true,labelfont=it]{subcaption}
\usepackage{colortbl}
\usepackage{enumitem}
\usepackage{tabularx}
\usepackage{graphicx}
\usepackage{booktabs}
\usepackage{array}
\usepackage{tcolorbox}
\usepackage[table]{xcolor}
\theoremstyle{definition}

\usepackage{booktabs}

\newcommand{\paratitle}[1]{\vspace{0.8ex}\noindent \textbf{#1}}

\newcommand{\our}{GenAIR}
\newcommand{\ours}{GenAIR\ }

\AtBeginDocument{%
  }

\copyrightyear{2026}
\acmYear{2026}
\setcopyright{cc}
\setcctype{by}
\acmConference[WWW '26]{Proceedings of the ACM Web Conference 2026}{April 13--17, 2026}{Dubai, United Arab Emirates}
\acmBooktitle{Proceedings of the ACM Web Conference 2026 (WWW '26), April 13--17, 2026, Dubai, United Arab Emirates}
\acmDOI{10.1145/3774904.3792587}
\acmISBN{979-8-4007-2307-0/2026/04}

\begin{document}

\title{Generative Archetype-Grounded Item Representations for Sequential Recommendation}

\author{Yifan Li}
\orcid{0009-0000-2414-367X}
\affiliation{%
  \institution{The Chinese University of Hong Kong}
  \city{Hong Kong SAR}
  \country{China}
}
\email{yfli24@cse.cuhk.edu.hk}

\author{Jiahong Liu}
\affiliation{%
  \institution{The Chinese University of Hong Kong}
   \city{Hong Kong SAR}
  \country{China}
  }
\email{jiahong.liu21@gmail.com}

\author{Xinni Zhang}
\affiliation{%
  \institution{The Chinese University of Hong Kong}
   \city{Hong Kong SAR}
  \country{China}
}
\email{xnzhang23@cse.cuhk.edu.hk}

\author{Hao Chen}
\affiliation{%
  \institution{The Chinese University of Hong Kong}
   \city{Hong Kong SAR}
  \country{China}
  }
\email{chenhao@cse.cuhk.edu.hk}

\author{Yankai Chen}
\affiliation{%
  \institution{McGill University}
   \city{Montreal}
  \country{Canada}
  }
\email{yankaichen@acm.org}

\author{Wenhao Yu}
\affiliation{%
  \institution{The Chinese University of Hong Kong}
   \city{Hong Kong SAR}
  \country{China}
  }
\email{whyu24@cse.cuhk.edu.hk}

\author{Jianting Chen}
\affiliation{%
  \institution{Tongji University}
   \city{Shanghai}
  \country{China}
  }
\email{tj_chenjt@tongji.edu.cn}

\author{Irwin King}
\affiliation{%
  \institution{The Chinese University of Hong Kong}
   \city{Hong Kong SAR}
  \country{China}
  }
\email{king@cse.cuhk.edu.hk}

\renewcommand{\shortauthors}{Yifan Li et al.}

\begin{abstract}

Sequential recommendation aims to predict users’ next interaction with items by analyzing their historical behavior. 
However, the limited quality of item representations remains a critical bottleneck.
While pre-trained large language models (LLMs) can provide rich semantic representations, existing approaches only rely on static encoding of fixed attributes, overlooking the crucial role of target audiences in defining item identity. Moreover, the semantic space struggles to reflect actual user behavior, resulting in a significant gap between semantic representations and behavioral patterns.
To address these limitations, we propose \textbf{GenAIR}, a general framework that empowers sequential recommendation with Generative Archetype-grounded Item Representations. 
Specifically, we first leverage an LLM to analyze item metadata and infer textual description of the Archetype, which represents the conceptual profile of the item’s ideal target audience. We then extract the corresponding embeddings in a single forward pass.
Further, to ground these generative archetypes in real-world behavior, we introduce a behavioral calibration objective,  which explicitly incorporates behavioral signals from actual interactions. This objective adjusts the structure of the embedding space to reflect empirical patterns.
GenAIR enables seamless integration with most existing models while maintaining high efficiency.
Comprehensive experiments conducted on three real-world datasets demonstrate that  GenAIR significantly improves the performance of various sequential recommendation models and consistently outperforms state-of-the-art baseline approaches.
Implementation codes are available at \url{https://github.com/AI-Santiago/GenAIR}.

\end{abstract}



\maketitle

\section{Introduction}

Recommendation systems are integral to digital experiences, as they shape how users navigate the vast array of content and products across online services ~\cite{he2017neural, DBLP:conf/kdd/0001TS0K24,  chen2023star, DBLP:conf/cikm/QiuZ0CLDK24, DBLP:conf/aaai/Zhang00FK24,wu2025ip2,zhang2023beyond,zhang2025survey}.
A prominent class of these systems is based on sequential recommendation,  which predicts users' next interaction with items based on their historical behavior.
In recent years, researchers have made considerable advances in  neural architectures~\cite{zhang2024wukong, hidasi2015session, tang2018personalized, kang2018self, sun2019bert4rec, liu2023diffusion, yang2023generate}, which learn  user embeddings by modeling the interactions between item representations.
Despite their success, these methods fundamentally operate on ID-based item representations, which are optimized through next-item prediction. 
Due to real-world data sparsity and imbalance issues,
these approaches suffer from limited  representational quality and generalization capability~\cite{zhao2023embedding, cao2022contrastive, xu2024rethinking}.
This limitation hinders the understanding of item-level signals and remains a critical bottleneck in fully modeling user intentions and profiles.

With the rise of large language models (LLMs), known for their robust knowledge representation and generative capabilities, they have 
provided new semantic perspectives for addressing this issue.
Recent studies~\cite{hu2024enhancing,liu2024practice, liu2024llmesr} have explored the utilization of LLMs to enhance item representations by encoding item textual metadata (e.g., names, brands, descriptions) into semantically rich embeddings. 
These representations inherit world knowledge from a large-scale pretraining process, enabling more generalized and informative item representations.  
Moreover, it naturally raises a key question: \textit{how to effectively obtain semantic embeddings and align them with the recommendation objectives?}
%最终目的是回答这个问题，而不是解决现有方法的limitation

Existing methods have proposed different representation learning strategies~\cite{harte2023leveraging, liu2024practice, liu2025llmemb, ren2024representation, hu2024enhancing,liu2024llmesr, li2024calrec}, aiming to optimize the semantic embedding space and enable integration with behavioral knowledge (e.g., item co-occurrences, clicks, purchases observed in real-world recommender systems). 
These include (\romannumeral 1)  \textbf{Text-based tuning}~\cite{harte2023leveraging, liu2024practice, liu2025llmemb, li2024calrec}, which fine-tunes the LLM with specific targets in textual form, designed to enhance its understanding of recommendations and modify the distribution of its semantic embedding through additional post-training.
(\romannumeral 2)  \textbf{Behavior-based tuning}~\cite{ren2024representation, hu2024enhancing,liu2024llmesr}, which bridges LLM semantic space with behavior space by matching them to embeddings from  recommendation models (trained either separately or jointly).
These methods enforce the two embedding spaces to be closer.

\begin{figure}[t]
    \centering
    \includegraphics[width=\linewidth, trim=26 150 16 0, clip]{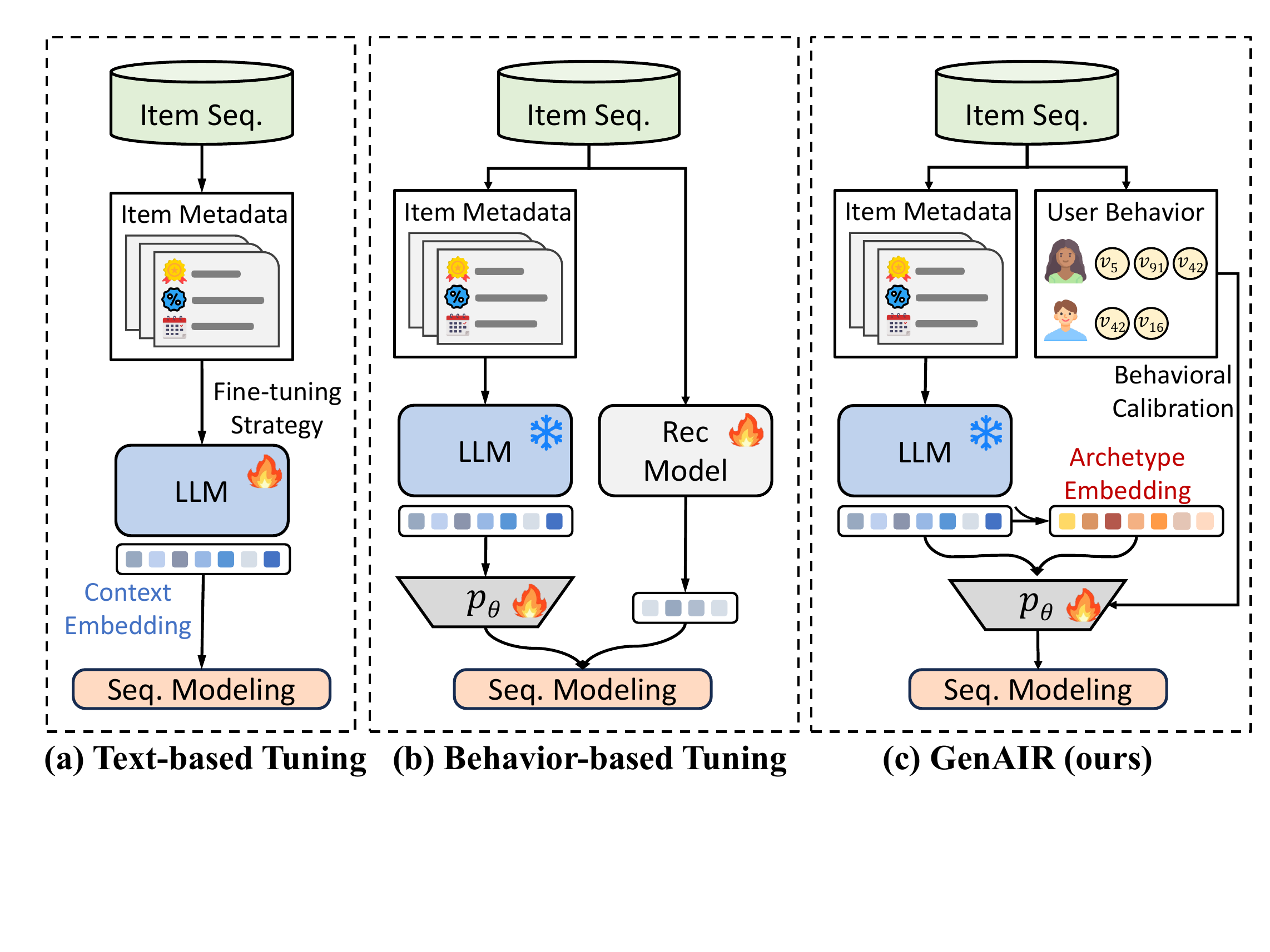}
\vspace{-15pt}
    \caption{
A brief comparison of different representation enhancement strategies, where $p_{\theta}$ denotes a projector module. The enhanced item representations are then fed into sequential recommendation models.
    }
    \label{fig:compare}
    \vspace{-20pt}
\end{figure}

Despite their promise, a critical analysis reveals three fundamental limitations:
(\romannumeral 1) \textbf{Neglect of Behavioral Information.} Text-based tuning methods (Figure~\ref{fig:compare}(a)) rely on static contextual text, which inherently neglects the behavioral patterns crucial for accurately modeling user interactions.
 In addition, such tuning methods demand substantial computational resources and exhibits limited efficiency. 
(\romannumeral 2) \textbf{Representational Mismatch.} As shown in Figure~\ref{fig:compare}(b), behavior-based tuning approaches
aim to bridge the semantic and behavioral spaces by aligning with ID embeddings from a separate recommendation model. However, ID embeddings suffer from data sparsity issues, and a significant modality gap between semantic and ID embeddings further limits their effectiveness as supervision signals~\cite{li2025g, zhao2023embedding, jang2020cities, kim2023melt}. 
(\romannumeral 3) \textbf{ Underutilized Generative Power.} A profound limitation shared by both paradigms is their failure to leverage the core strength of modern LLMs. 
Existing frameworks relegate the LLM to the passive role of a feature extractor or a static text encoder~\cite{harte2023leveraging, liu2025llmemb}, neglecting their generative and reasoning abilities~\cite{guo2025deepseek, jaech2024openai}.  
The potential to harness these capabilities to interpret targeted user intent and profiles remains largely unexplored.

%ours
\paratitle{Presented Work.} Motivated by these challenges,
in this paper, we propose \textbf{\our}, a general framework  that empowers sequential recommendation with \textbf{Gen}erative \textbf{A}rchetype-grounded \textbf{I}tem \textbf{R}epresentations.
We  define an \textit{Archetype} as a conceptual representation, which embodies the hypothetical user groups whose preferences align most strongly with the item. This approach is grounded in the STP (Segmentation, Targeting, Positioning) framework~\cite{kotler1969broadening}, which holds that an item's identity is shaped not just by its attributes, but by its target audience. 
To this end, as illustrated in Figure~\ref{fig:compare}(c), we first leverage an LLM to analyze item metadata and generate latent user archetypes.
We then obtain semantic item embeddings for these archetypes (referred as archetype embeddings), capturing latent behavioral preferences at the semantic level beyond static attributes.
However, while utilizing LLM's world knowledge reveals an item's potential user profile, the real user group often requires actual interaction behavior  to emerge. 
Therefore, to ensure these representations align closely with real user behavior, we introduce a new training objective. 
Specifically, we capture the collective characteristics of items and their actual audiences based on their association in real-world behavior, and subsequently introduce a behavioral calibration objective that grounds these generative  representations in real interaction patterns.
Our framework offers seamless integration and compatibility with existing sequential recommendation methods, and maintains high computational efficiency throughout the training process and introduces no additional overhead during inference.
Overall, our contributions can be summarized as follows:
\begin{itemize}[leftmargin=*]
    \item We propose archetype-grounded item representations, where items are characterized through generative modeling of their target audiences rather than static attributes alone.
    \item We present a general framework that aligns LLM-generated semantic representations with user interactions through a behavioral calibration objective.
    \item We validate the effectiveness of \ours through extensive experiments across three datasets, where it consistently outperforms state-of-the-art baselines, demonstrating its superior performance and practical applicability.
\end{itemize}

\section{Preliminary}

\noindent\textbf{Problem Statement.}
Sequential recommendation aims to predict the next item a user would interact with, given the historical interaction sequence ~\cite{fang2020deep}. 
Let $\mathcal{V}$ denote the universal set of items, where the item $i$ is represented as $v_i \in \mathcal{V}$. The history of user interactions is ordered chronologically and formalized as the sequence $\mathcal{Q}$.
The task is to predict the next item $v_{N+1}$ by solving:
\begin{equation}
    v_{N+1}^* = \mathop{\mathrm{arg\,max}}\limits_{v_i \in \mathcal{V}} \, P \left( v_{N+1} = v_i \mid \mathcal{Q} \right).
\end{equation}

\noindent\textbf{Model Formulation.}
Most sequential recommendation models follow an embedding-sequence framework. 
First, the item $v_i$ is mapped to a dense embedding: 
\begin{equation}
\mathbf{e}_i = {\mathrm {Emb}}(v_i), 
\end{equation}
where ${\mathrm {Emb}}(\cdot)$ is the embedding function, and $\mathbf{e}_i \in \mathbb{R}^d$ captures item relationships in a high-dimensional space. 
Then the sequence model extracts and produces a user representation: 
\begin{equation}
\mathbf{u} = {\mathrm {Seq}}(\{\mathbf{e}_1, \mathbf{e}_2, \dots, \mathbf{e}_N\}) \in \mathbb{R}^d, 
\end{equation}
where ${\mathrm Seq}(\cdot)$ is the  backbone recommendation model.
The recommendation probability for the item $v_i$ is computed as: 
\begin{equation}
P(v_{N+1}=v_i|\mathcal{Q}) = \mathbf{u}^T \mathbf{e}_i.
\end{equation}
However, the existing embedding function ${\rm Emb}(\cdot)$ often relies solely on item IDs.
Our goal is to leverage LLMs to enhance the embedding function to achieve better representations and facilitate their integration into various backbone recommendation models.

\section{Methodology}

\subsection{Archetype Generation}
\label{sec:method1}

\subsubsection{Archetype Instantiation}
\begin{figure*}[t]
    \centering
    \includegraphics[width=\linewidth, trim=0 25 0 90, clip]{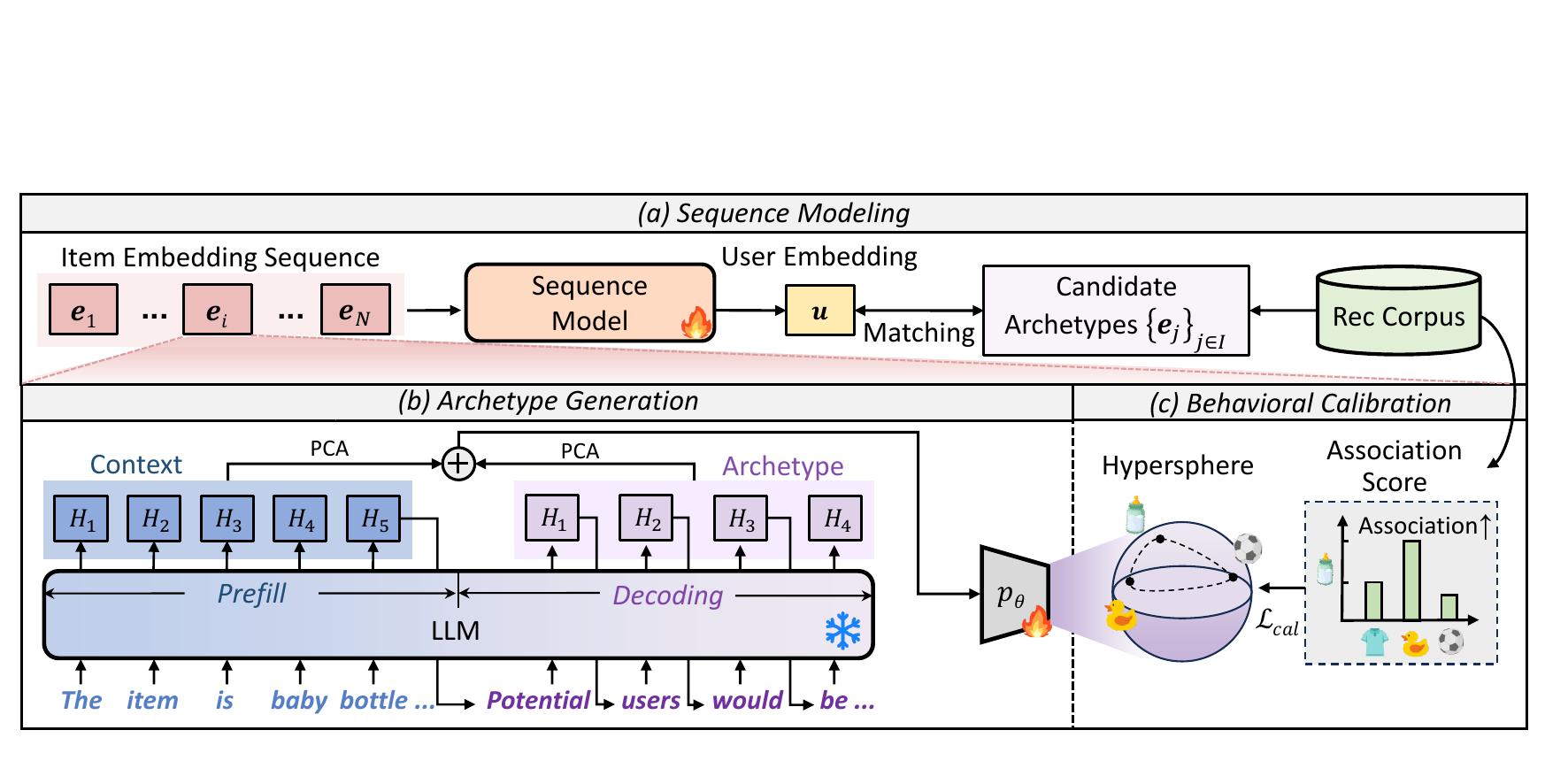}
        \caption{The overview of \our. (a) Sequence Modeling: The item representations are organized into sequences and processed by the sequence model for sequential recommendation. (b) Archetype Generation: Partial words are shown for illustration, and for simplicity, we assume that each word corresponds to a token, with $H_i$ denoting the hidden state of the LLM's last layer for the $i$-th token. (c) Behavioral Calibration:  The projector $p_\theta$ is optimized using an enhanced association-based objective.} 
        \label{fig:model}
        \vspace{-7pt}
\end{figure*}

The key to recommendation lies in establishing consistent patterns between user preferences and item characteristics, as shown in Figure~\ref{fig:model}(a). This consistency is typically learned from implicit signals within historical user-item interaction data, such as clicks, purchases, and ratings. However, such signals are inherently retrospective, primarily reflecting the result of complex user-item relationships while overlooking their underlying causes.
Drawing on the STP (Segmentation, Targeting, Positioning) framework~\cite{kotler1969broadening}, we posit that every item inherently embeds an idealized target user profile throughout its lifecycle, from conceptual design and functional implementation to its marketing. This profile guides strategic decisions and ultimately shapes its market positioning and appeal. As it represents the provider's inference based on their knowledge and assumptions rather than a direct description of any specific user, we define it as an  \textbf{archetype}.
For item $i$, archetype $A_i$ serves as a rich and corresponding representation of its target users, providing a foundation for anchoring the item within the nuanced space of human preferences.

However, this concept presents two computational challenges. First, these archetypes are often internal inferences of item providers and are not accessible to recommendation platforms. Second, existing recommendation models lack the capability to infer target user profiles directly from item attributes. To address these challenges, we propose a systematic approach to reconstruct latent user archetypes from publicly available item metadata. Specifically, an item’s design intent is implicitly encoded in its metadata, from which we can derive latent user archetypes via semantic reasoning.
 This process relies on deep semantic understanding and commonsense reasoning capabilities, making recent LLMs well-suited~\cite{touvron2023llama, qwen2.5}.
Therefore,  we use an LLM to instantiate the abstract concepts of the archetypes into concrete textual forms, providing rich features for downstream tasks while ensuring consistency in descriptive paradigms across items. 

For each item $i$ in the catalog $\mathcal{I}$, we organize the item's metadata attributes (e.g., \texttt{name}, \texttt{brand}, \texttt{category}) into a structured text input $C_i$. We then fill this contextual information into a prompt template shown in Table~\ref{tab:prompt}, which is chosen for its simplicity. 
This template guides and constrains the LLM's reasoning process, instructing the model to understand the item metadata and generate descriptive content from the user archetype perspective. By calling an LLM, we obtain a textual description of the archetype $A_i$ for item $i$:
\begin{equation}
A_i = \mathrm{LLM}(\mathrm{PromptTemplate}(C_i)).
\end{equation}
This generation step is performed for each item in the catalog $ \mathcal{I}$, yielding a collection of archetypes $\{A_i\}_{i \in \mathcal{I}}$. A detailed example can be found in Appendix~\ref{appendix:example}.

\begin{table}[t]
\centering
\caption{Prompt Template.}
\vspace{-12pt}
\label{tab:prompt}
\begin{tcolorbox}[
    colback=gray!5,
    colframe=gray,
    width=1.0\linewidth,
    sharp corners,
    title={\textbf{Prompt Template}},
    fontupper=\normalsize,   % 指定正文字号
    fonttitle=\normalsize\bfseries % 指定标题字号
]
You are a helpful recommendation expert.\\
Based on the following attributes, identify the types of users this item would appeal to: \\
The item name is \texttt{<item\_name>}. \\
The item brand is \texttt{<item\_brand>}. \\
The item date is \texttt{<item\_date>}. \\
The item feature is \texttt{<item\_feature>}. \\
The item price is \texttt{<item\_price>}. \\
The item rating is \texttt{<item\_rating>}. \\
The item description is \texttt{<item\_description>}. \\

\textbf{What potential users align best with this item?}  
\end{tcolorbox}
\vspace{-21pt}
\end{table}

% \vspace{-15pt}
\begin{figure}[t]
\centering
    \begin{subfigure}[b]{0.49\linewidth}
    \centering
    \includegraphics[width=\linewidth]{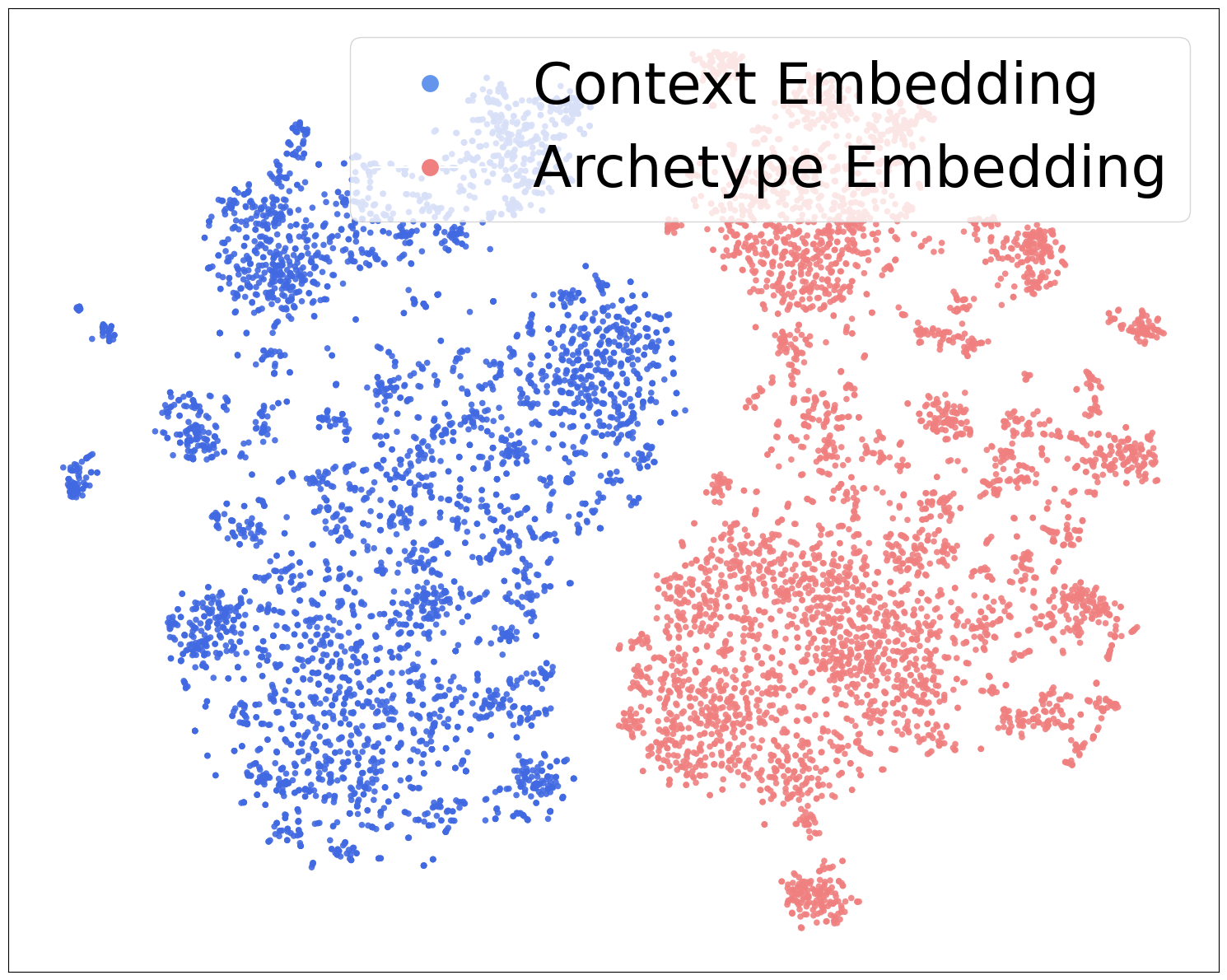}
      \subcaption{Amazon Fashion}

    % \end{subfigure}
    \end{subfigure}\hspace{-0.1cm}
    \begin{subfigure}[b]{0.49\linewidth}
    \centering
    \includegraphics[width=\linewidth]{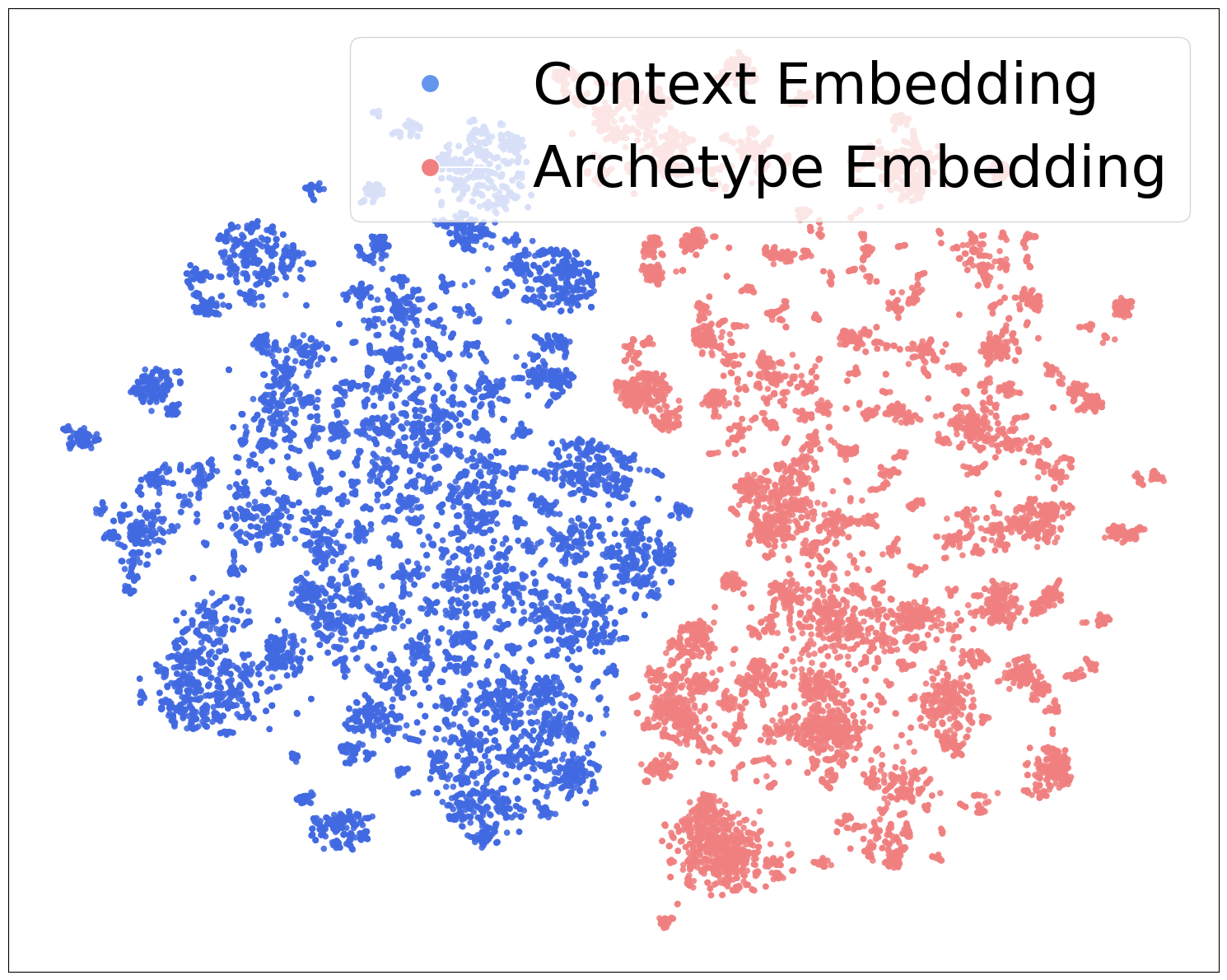}
    \subcaption{Yelp}

    \end{subfigure}

        \vspace{-5pt}
        \caption{Visualization of representation space.}
        \vspace{-21pt}
\label{fig:space}
\end{figure}

\subsubsection{Archetype Embeddings}

After generating archetypes from the metadata context, the next step is to extract these rich language-based representations into unified numerical embeddings. A key insight of our approach is that both the contextual text and the generated archetype text can be embedded through a single forward pass of an autoregressive LLM. We obtain the embeddings from the hidden states of the LLM's last transformer layer, as shown in Figure~\ref{fig:model}(b). 
This ensures efficient embedding access while keeping them in a shared, semantically coherent space.

This process leverages the two phases of a standard generative LLM invocation. First, during the \textit{prefill} phase, the LLM processes an input prompt containing the serialized metadata $C_i$ and computes a sequence of final hidden states, denoted as $\mathbf{H}_i^{\mathrm{prefill}} \in \mathbb{R}^{n_p \times d_{\mathrm{LLM}}}$, where $n_{p}$ represents the number of input tokens and $d_{\mathrm{LLM}}$ denotes the hidden dimension of LLM. Immediately following, during the \textit{decoding} phase, the model generates archetype text $A_i$ while producing a corresponding sequence of hidden states. Each generated token is one-to-one with a hidden state $\mathbf{H}_i^{\mathrm{decoding}} \in \mathbb{R}^{n_d \times d_{\mathrm{LLM}}}$, where $n_{d}$ represents the number of generated tokens.
To derive a fixed-size embedding, we apply mean pooling along the token dimension ~\cite{behnamghaderllm2vec, muennighoff2024generative, tang2024pooling}:
\begin{equation}
    \mathbf{E}_i^{\mathrm{prefill}} = \frac{1}{n_p} \sum_{j=1}^{n_p} \mathbf{H}_{i, [j,:]}^{\mathrm{prefill}}, \quad \quad
    \mathbf{E}_i^{\mathrm{decoding}} = \frac{1}{n_d} \sum_{j=1}^{n_d} \mathbf{H}_{i, [j,:]}^{\mathrm{decoding}}.
\end{equation}
The \textit{prefill embedding} $\mathbf{E}_i^{\mathrm{prefill}}$ reflects the item's factual attributes, while the \textit{decoding embedding} $\mathbf{E}_i^{\mathrm{decoding}}$ captures the inferred description of the hypothetical user. 
Our initial finding  (obtained from LLama 2-7B-Chat ~\cite{touvron2023llama} on different datasets) shows the t-SNE visualization space in Figure ~\ref{fig:space}, where two  components are naturally distinguished, capturing diversity and contributing aspects.

To  bridge the large dimensional gap between LLM embeddings and recommendation model embeddings, we employ a standard projection approach. Following prior  work~\cite{liu2025llmemb, liu2024llmesr}, we begin by applying principal component analysis (PCA) ~\cite{pearson1901liii} to each embedding for signal-to-noise separation, retaining principal components sufficient to explain 95\% of the variance.
Let $\hat{\mathbf{E}}_i^{\mathrm{prefill}}$ and $\hat{\mathbf{E}}_i^{\mathrm{decoding}}$ denote the dimension-reduced embeddings. These are concatenated to construct the final embedding $\mathbf{E}_i$:
\begin{equation}
\label{eq:concat}
\mathbf{E}_i = \mathrm{Concat}(\hat{\mathbf{E}}_i^{\mathrm{prefill}}, \hat{\mathbf{E}}_i^{\mathrm{decoding}}).
\end{equation}
This representation is grounded in the item-specific features while simultaneously being enriched by the user-centric perspective of its generated archetype.
We freeze these representations to prevent semantic degradation.
Next, for better alignment and compatibility with recommendation models, we apply a trainable  projector module $p_{\theta}$ to map the embedding into the  model’s latent space. In practice, $p_{\theta}$  is implemented as a Multi-Layer Perceptron (MLP).
\begin{equation}
\label{eq:p}
    \mathbf{e}_{i} =  p_{\theta}({\mathbf{E}_{i}}).
\end{equation}
These embeddings will then serve as the semantically aware item representations for the subsequent stages of the recommendation model and are used to match with user embeddings.

\subsection{Behavioral Calibration}
In Section~\ref{sec:method1}, we obtain the archetype for each item using an LLM, constructing semantic item representations.
These embedded vectors serve as powerful initializations, capturing the expected characteristics and target audiences of items. 
However, purely semantic representations exist in isolation from real-world user behavior,  which often exhibits structured behavioral patterns far more complex than semantic labels. 
For instance, users tend to purchase baby bottles and toys together, yet rarely buy baby bottles and soccer balls simultaneously.
These associations stem from empirical behavior rather than being determined solely by semantics.
Therefore, to capture the actual engagement patterns reflecting group acceptance in actual interactions, we propose behavioral calibration, a mechanism that refines the initial semantic embedding space using association-based behavioral signals, as shown in Figure~\ref{fig:model}(c).

Following prior work in representation learning ~\cite{xu2018spherical, mettes2019hyperspherical}, we  project all item embeddings onto the unit hypersphere  
by applying $\ell_2$ normalization: $e_i \leftarrow e_i / |e_i|_2$. 
This creates a common manifold, where angular distance can serve as a meaningful metric for measuring inter-item relationships.
A common objective for such representations is to promote uniformity, encouraging embeddings to spread across the space to maximize entropy and expressive capacity. This can be framed as a uniformity objective centered around a distance-based Gaussian kernel~\cite{wang2020understanding}:
\begin{equation}
G_{t}(e_i, e_j) = t \| \mathbf{e}_i - \mathbf{e}_j \|_2^2 ,
\end{equation}
where $t$ is a fixed temperature parameter.
However, while promoting diversity,  it operates on a flawed assumption that all items should be pushed apart equally, which fundamentally misaligns with the relational structure of user preferences.

To construct a representation space aligned with actual behavior, we argue that repulsion requires modulation based on empirical engagement interactions, such as co-clicks, co-purchases, or other forms of behavioral co-occurrence.  
To achieve this, we first quantify the behavioral signal from interaction logs, and define it with the inter-item co-occurrence statistics
$C(i, j)$, which counts how frequently two items co-appear within the same user context, capturing implicit behavioral proximity. To ensure these statistics are stable and comparable across pairs with varying popularity, we define a log-scaled formulation to calculate the association score:
\begin{equation}
S(i, j) =  \frac{\log(1 + C(i, j))}{\log(1 + C(i, j)_{\max})},
\label{eq:weighting}
\end{equation}
where $ C(i, j)_{\max}$ is the maximum co-occurrence count across all item pairs. 
The association score $S(i, j)$ is then used to construct the behavioral regulator $w(i, j)$, a data-driven exponential decay function that controls the repulsion for each specific pair $(i, j)$:
\begin{equation}
w(i, j) = \exp\big(-\gamma \cdot S(i, j)\big), \quad \gamma > 0,
\end{equation}
where $\gamma$ is a positive hyperparameter that controls the decay rate of the repulsive force between behavioral associations.
We thus adaptively regulate the repulsive force, implicitly achieving calibration through differential repulsion.
\begin{align}
K_t(i, j) &= w(i, j) \cdot \exp\!\left(-t \lVert \mathbf{e}_i - \mathbf{e}_j \rVert_2^2 \right), \label{eq:wkernel1} \\
  &= \exp\left(-\gamma \cdot S(i, j) + \beta \mathbf{e}_i^T \mathbf{e}_j  \right), \label{eq:wkernel3}
\end{align}
where $\beta = 2t$ \footnote{The constant term from the squared distance formulation is absorbed.}, and the reformulation from \eqref{eq:wkernel1} to \eqref{eq:wkernel3} exploits the identity between squared Euclidean distance and dot product under $\ell_2$-normalized embeddings, aligning the objective directly with the similarity metric widely adopted in recommender systems.
The calibration objective is formulated by minimizing the expected kernel function value. The loss $\mathcal{L}_{\text{cal}}$ is defined as:
\begin{equation}
\mathcal{L}_{\mathrm{cal}} = \mathbb{E}_{(\mathbf{e}_i, \mathbf{e}_j) \sim p_{\mathrm{data}}} \left[ K_t(i, j) \right].
\label{eq_cal}
\end{equation}
In practice, we approximate the expectation over a mini-batch $\mathcal{B}$. Further, to build a clear intuition for how $\mathcal{L}_{{cal}}$ structures the embedding space, we analyze its gradient and reveal that it shapes the space through regulated inner‑item interactions.

\begin{proposition}[Gradient as a Weighted Force]
\label{prop:force-law}
The gradient  with respect to an embedding $e_i$ is a weighted sum of repulsive forces from all other embeddings, e.g., $e_j$, in the mini-batch:
\begin{equation}
\frac{\partial\mathcal{L}_{\mathrm{cal}}}{\partial \mathbf{e}_i} 
= \frac{\beta}{|\mathcal B|} \sum_{j : (i,j)\in\mathcal{B}} 
w(i,j)  \exp({\beta \mathbf{e}_i^T \mathbf{e}_j} )\mathbf{e}_j.
\end{equation}
\end{proposition}
Proof of Proposition~\ref{prop:force-law}  is in Appendix~\ref{appendix:proof1}.
The magnitude not only increases with the dot product but is also scaled  by $w(i,j)$, thereby injecting prior structure into the embedding space.

\subsection{Model Training and Inference}

The learned embeddings $\{ \mathbf{e}_i \}$ are model-agnostic, which can be integrated into most sequential recommendation models, such as GRU4Rec~\cite{hidasi2015session},  BERT4Rec~\cite{sun2019bert4rec}, and SASRec~\cite{kang2018self}.
\subsubsection{Training}
During the training process of the recommendation model,  
we construct sequences based on the interaction history with $\{ \mathbf{e}_i \}$, then feed them into the sequence model to obtain the user embedding $\mathbf{u}$.
The probability of recommending an item $j$ to user $u$ is computed as the dot product:
\begin{equation}
\label{eq_dot}
P(v_{n_u+1}=v_j \mid v_{1:n_u}) = \mathbf{e}_j^T \mathbf{u}.
\end{equation}

The pairwise ranking loss for the  recommendation model is calculated as:
\begin{equation}
\label{eq_rank}
\mathcal{L}_{\mathrm{rank}} = -\sum_{u \in \mathcal{U}} \sum_{k=1}^{n_u} \log \sigma \left( P(v^{+}_{k+1} \mid v_{1:k}) - P(v^{-}_{k+1} \mid v_{1:k}) \right),
\end{equation}
where $n_u$ denotes the sequence length, $v^{+}_{k+1}$ is the ground-truth next item, and $v^{-}_{k+1}$ is a sampled negative item. 
Here, the ranking loss formulation varies depending on the sequence model, such as adopting a sequence-to-one pairwise loss for models like GRU4Rec ~\cite{hidasi2015session}.

The entire model is trained end-to-end by minimizing a composite objective function, which is formulated as:
\begin{equation}
\label{eq_sum}
\begin{aligned}
\mathcal{L}_{\mathrm{}} = \mathcal{L}_{\mathrm{rank}} + \alpha \mathcal{L}_{\mathrm{cal}} ,
\end{aligned}
\end{equation}
where the hyper-parameter $\alpha$ adjusts the contributions of $\mathcal{L}_{\mathrm{cal}}$.
\subsubsection{Inference}
We obtain the embedding by Equation~\ref{eq:p}. 
Since the representations of items can be cached in advance, 
LLM calls are avoided during the inference phase, 
which would not introduce any additional inference cost for the recommendation model.
We conclude the above process in Algorithm~\ref{algorithm} for clarity.

\begin{algorithm}[t]
\caption{Training and Inference}
\label{algorithm}
\begin{algorithmic}[1]
\Procedure{Train}{$\mathcal{D}_{\mathrm{train}}, \mathbf{E}^{\mathrm{prefill}}, \mathbf{E}^{\mathrm{decoding}}, \lambda, \eta$}
    \State Initialize model parameters $\theta$;
    \State $\mathbf{E}_{\mathrm{fused}} \gets \mathrm{Fuse}(\mathbf{E}^{\mathrm{prefill}}, \mathbf{E}^{\mathrm{decoding}})$;
    % \Comment{Eq.~\eqref{eq:concat} and \eqref{eq:p}}    
    \For{$e \gets 1$ to $N_{\mathrm{epochs}}$}
        \For{each mini-batch $(\mathcal{S}, \mathcal{Y}^+, \mathcal{Y}^-) \subset \mathcal{D}_{\mathrm{train}}$}
            \State $\{\mathbf{h}_u\} \gets \mathrm{Seq}(\mathcal{S}, \mathbf{E}_{\mathrm{fused}}; \theta)$; 
            % \Comment{Backbone}

            \State $\mathcal{L}_{\mathrm{cal}} \gets \mathrm{ComputeCalibrationLoss}(\{\mathbf{h}_u\})$; 
            % \Comment{Eq.~\eqref{eq_cal}}
            
            \State $\mathcal{L}_{\mathrm{rank}} \gets \mathrm{ComputeRankingLoss}(\{\mathbf{h}_u\}, \mathcal{Y}^+, \mathcal{Y}^-)$; 
            % \Comment{Eq.~\eqref{eq_rank}}
            \State $\mathcal{L} \gets \mathcal{L}_{\mathrm{rank}} + \alpha \mathcal{L}_{\mathrm{cal}}$; 
            % \Comment{Eq.~\eqref{eq_sum}}

        \State $\theta \gets \theta - \eta \nabla_{\theta} \mathcal{L}$;
        % \Comment{Update model parameters}
        
        \EndFor
    \EndFor
    \State \Return Trained parameters $\theta$.
\EndProcedure
\Statex
\Procedure{Recommend}{$\mathcal{S}_u, \mathcal{I}_{\mathrm{candidates}}, \theta, \mathbf{E}_{\mathrm{fused}}$}
    \State $\mathbf{h}_u, \{\mathbf{v}_i\}_{i \in \mathcal{I}} \gets \mathrm{Seq}(\mathcal{S}_u, \mathbf{E}_{\mathrm{fused}}; \theta)$; 
    \State $Scores \gets \{\}$;
    \For{each item $i \in \mathcal{I}_{\mathrm{candidates}}$}
        \State $s_i \gets \mathbf{h}_u^\top \mathbf{v}_i$; 
        % \Comment{Eq.~\eqref{eq_dot}}
        
        \State $Scores[i] \gets s_i$;
    \EndFor
    \State $\mathcal{R}_u \gets \mathrm{SortByValue}(Scores, \mathrm{descending=True})$;
    \State \Return Top-$K$ items from $\mathcal{R}_u$.
\EndProcedure
\end{algorithmic}
\end{algorithm}
\section{Experiments}
\label{experiment}

\begin{table*}[t]
\caption{
Overall performance comparisons between competing baselines and our \ours across different backbones on three datasets. 
Bolded values indicate the best results, showing statistically significant improvements ($p < 0.05$, two-sided t-test) compared to the second-best (underlined) baseline.
}
\centering
\tabcolsep=0.21cm   
\resizebox{0.97\textwidth}{!}{
\begin{tabular}{l|c|c|ccc|ccccc|c>{\columncolor{gray!10}}c}
\toprule

 \multicolumn{3}{c|}{\textbf{General  Setting}} &\multicolumn{3}{c|}{\textbf{Traditional Method}} &\multicolumn{5}{c|}{\textbf{Language-based Method}} & \multicolumn{2}{c}{\textbf{Ours}} \\

\midrule

\textbf{Dataset} & \textbf{Backbone} & \textbf{Metric}  & Base   & CITIES  & MELT   & RLMRec  &LLMInit &LLM-ESR  &LLMEmb &Alphafuse &\textbf{GenAIR} & Imprv.\\ 

\midrule

\multirow{6}{*}{\textbf{Yelp}} 
& \multirow{2}{*}{\textbf{GRU4Rec}}  & HR@10 & 0.4879 & 0.4898 & 0.4985  & 0.4824 & 0.5323 & 0.5592 & 0.5592 &\underline{0.5636}     & \textbf{0.5775} &+2.46\%\\
&  & NDCG@10 & 0.2751 & 0.2749 & 0.2825  & 0.2765 & 0.3132 & 0.3300 & 0.3301  & \underline{0.3380}    & \textbf{0.3418} &+1.12\% \\

\cmidrule{2-13} 

& \multirow{2}{*}{\textbf{Bert4Rec}}  & HR@10 & 0.5307 & 0.5249 & 0.6206  & 0.5356 & 0.5518 & 0.5886 & 0.6181  &\underline{0.6565}     & \textbf{0.6912} &+5.29\% \\
&  & NDCG@10 & 0.3035 & 0.3015 & 0.3770  & 0.3111 & 0.3217 & 0.3551 & 0.3907  & \underline{0.4111}    & \textbf{0.4390} &+6.79\% \\

\cmidrule{2-13} 

& \multirow{2}{*}{\textbf{SASRec}}  & HR@10 & 0.5940 & 0.5828 & 0.6257  & 0.5929 & 0.6302 & 0.5935 & 0.6488  &\underline{0.6658}     & \textbf{0.6797} &+2.09\% \\
& & NDCG@10 & 0.3597 & 0.3540 & 0.3791  & 0.3583 & 0.3859 & 0.3550 & 0.4139  &\underline{0.4171}     & \textbf{0.4302} &+3.14\% \\

\midrule

\multirow{6}{*}{\textbf{Beauty}} 
& \multirow{2}{*}{\textbf{GRU4Rec}}  & HR@10 & 0.3683 & 0.2456 & 0.3702  & 0.3690 & 0.4333 & 0.4765 & 0.4507  & \underline{0.4788}    & \textbf{0.5121} &+6.95\% \\
 & & NDCG@10 & 0.2276 & 0.1400 & 0.2161 & 0.2295  & 0.2588 & 0.2992 & 0.2863  & \underline{0.3046}    & \textbf{0.3338} &+9.59\% \\

\cmidrule{2-13} 

& \multirow{2}{*}{\textbf{Bert4Rec}}  & HR@10 & 0.3984 & 0.3961 & 0.4716  & 0.4019 & 0.4821 & 0.4790 & 0.4482  & \underline{0.5627}    & \textbf{0.5936} &+5.49\% \\
 & & NDCG@10 & 0.2367 & 0.2339 & 0.2965  & 0.2391 & 0.3039 & 0.2920 & 0.2663  & \underline{0.3747}    & \textbf{0.3961} &+5.71\% \\

\cmidrule{2-13} 

& \multirow{2}{*}{\textbf{SASRec}}  & HR@10 & 0.4388 & 0.2256 & 0.4334 & 0.4236 & 0.5263  & 0.4949 & 0.5036  & \underline{0.5585}    & \textbf{0.5939} &+6.34\% \\
 & & NDCG@10 & 0.3030 & 0.1413 & 0.2775 & 0.2866 & 0.3551  & 0.3056 & 0.3153  & \underline{0.3754}    & \textbf{0.4074} &+8.52\% \\

\midrule

\multirow{6}{*}{\textbf{Fashion}} 

& \multirow{2}{*}{\textbf{GRU4Rec}}  & HR@10 & 0.4798 & 0.4762 & 0.4884 & 0.4867 & 0.4959 & \underline{0.5514}  & 0.5414  & 0.5346    & \textbf{0.5762} &+4.50\% \\
 & & NDCG@10 & 0.3809 & 0.3743 & 0.3975 & 0.4065 & 0.3932 & \underline{0.4555}  & 0.4547  & 0.4375    & \textbf{0.4805} &+5.49\% \\

\cmidrule{2-13}  

& \multirow{2}{*}{\textbf{Bert4Rec}}  & HR@10 & 0.4668 & 0.4926 & 0.4897 & 0.4734 & 0.4886 & \underline{0.5432} & 0.5244   & 0.5289    & \textbf{0.5558} &+2.32\% \\
 & & NDCG@10 & 0.3613 & 0.4090 & 0.3810 & 0.4012 & 0.3610 & \underline{0.4421}  & 0.4238  & 0.4215    & \textbf{0.4463} &+0.95\% \\

\cmidrule{2-13} 

& \multirow{2}{*}{\textbf{SASRec}}  & HR@10 & 0.4956 & 0.4923 & 0.4875 & 0.5013 & 0.5169 & 0.5398 & \underline{0.5872}   &0.5632     & \textbf{0.6133} &+4.44\% \\
 & & NDCG@10 & 0.4429 & 0.4423 & 0.4150 & 0.4462 & 0.4138 & 0.4363  & \underline{0.4973}  & 0.4529    & \textbf{0.5207} &+4.71\% \\

\bottomrule
\end{tabular}
}

\label{tab:Overall}
\end{table*}

In this section, we conduct comprehensive experiments across different datasets to address the following research questions.

\begin{itemize}[leftmargin=*]
\item \textbf{RQ1:} 
How does the proposed \ours perform across different sequential recommendation models?

\item \textbf{RQ2:} How do different components and hyper-parameter settings affect the performance of \our?

\item \textbf{RQ3:} How do different base LLMs impact the performance?

\item \textbf{RQ4:} What are the computational costs of \ours during training and inference?
\item \textbf{RQ5:} How does \ours affect different item groups?

\end{itemize}

\subsection{Experimental Settings}
\subsubsection{Datasets}

We conduct experiments on three  real-world datasets for evaluation, namely Yelp, Amazon Fashion, and Amazon Beauty.
We strictly follow the previous studies ~\cite{kang2018self,liu2024llmesr} for preprocessing and data split.  
More details of the datasets  are in \textbf{Appendix}~\ref{appendix:datasets}.

        \vspace{-1pt}

\subsubsection{Backbones and Baselines.}
 To validate the generality, following common configurations~\cite{liu2024llmesr, liu2025llmemb}, we experiment on the following backbone models: GRU4Rec~\cite{hidasi2015session}, Bert4Rec ~\cite{sun2019bert4rec} and SASRec~\cite{kang2018self}. Further, to validate the effectiveness, we compare \ours with several representative methods. Our comparisons include traditional methods such as CITIES~\cite{jang2020cities} and MELT~\cite{kim2023melt}. Additionally, we benchmark against the most recent language-based approaches that incorporate semantic embeddings from LLMs, including RLMRec~\cite{ren2024representation}, LLMInit~\cite{harte2023leveraging, hu2024enhancing, qu2024elephant},  LLM-ESR~\cite{liu2024llmesr}, LLMEmb~\cite{liu2025llmemb}, and Alphafuse~\cite{DBLP:conf/sigir/Hu0LC0025}. 
More details can be found in \textbf{Appendix}~\ref{appendix:baselines}.

        \vspace{-1pt}

\subsubsection{Implementation Details}
We use LLama 2-7B-Chat ~\cite{touvron2023llama} as the foundation LLM for the main results.
And we fix the final embedding dimensionality to 128 for all methods.
For the baseline that utilizes an additional loss function, we use the optimal coefficients suggested in their original paper  ~\cite{liu2024llmesr, liu2025llmemb}.
For \our,  we set $t=2$ following ~\cite{wang2020understanding}, and search for the hyper-parameters $\gamma$ and $\alpha$ from $\{0, 1, 2, 3, 4, 5\}$ and $\{0.0001, 0.0005, 0.001, 0.005, 0.01, 0.05, 0.1\}$, respectively.
Optimization is performed using the Adam optimizer.
More detailed settings can be found in Appendix~\ref{appendix:implementation}.

        \vspace{-1pt}

\subsubsection{Evaluation Protocols and Metrics}
We evaluate the performance using two common metrics:
Hit Ratio (HR@10) and Normalized Discounted Cumulative Gain (NDCG@10). 
For the robustness of the results, we calculate the 
average results
obtained from three independent runs.

\begin{table}[t]
\caption{The ablation study on the Fashion dataset with different  models. The highest scores are in bold.}
        \vspace{-1pt}
\centering
\resizebox{1.0\linewidth}{!}{%
\begin{tabular}{l|c|cccc}
\toprule
\textbf{Model} & \textbf{Metric} & w/o Pref. & w/o Decod. & w/o Calib. & \textbf{\our} \\ 
\midrule
\multirow{2}{*}{GRU4Rec}   & HR@10   & 0.5621 &0.5636  & 0.5627 & \textbf{0.5762} \\ 
                          & NDCG@10 &0.4752  & 0.4730 & 0.4760 & \textbf{0.4805} \\ 
\midrule

\multirow{2}{*}{Bert4Rec}   & HR@10   & 0.5311 & 0.5372 & 0.5388 & \textbf{0.5558} \\ 
                          & NDCG@10 & 0.4355 & 0.4328 & 0.4402 & \textbf{0.4463} \\ 
\midrule

\multirow{2}{*}{SASRec}   & HR@10   & 0.5968 & 0.6010 & 0.6085 & \textbf{0.6133} \\ 
                          & NDCG@10 & 0.5069 & 0.5037 & 0.5171 & \textbf{0.5207} \\

\bottomrule
\end{tabular}%
}

\label{tab:exp_ablation}
\vspace{-10pt}
\end{table}

\begin{figure*}[t]
\centering
    \begin{subfigure}[b]{0.24\linewidth}
    \centering
      \includegraphics[width=\linewidth]{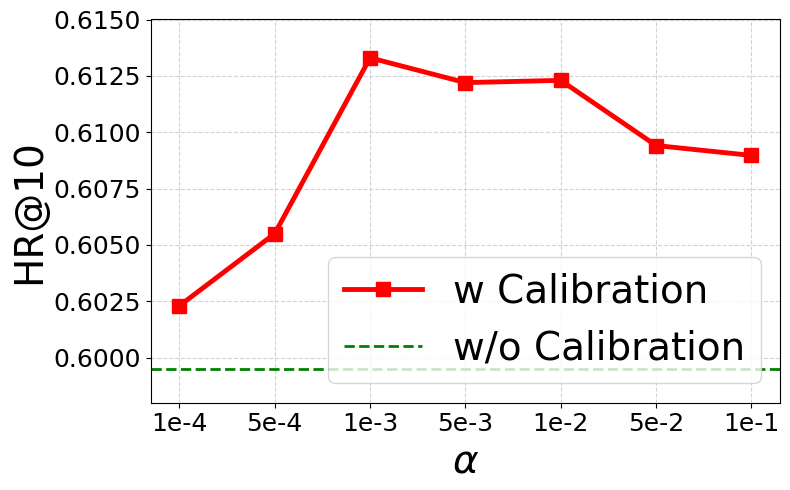}
      \subcaption{Impact of $\alpha$ on HR@10}
    \end{subfigure}
    \begin{subfigure}[b]{0.24\linewidth}
    \centering
    \includegraphics[width=\linewidth]{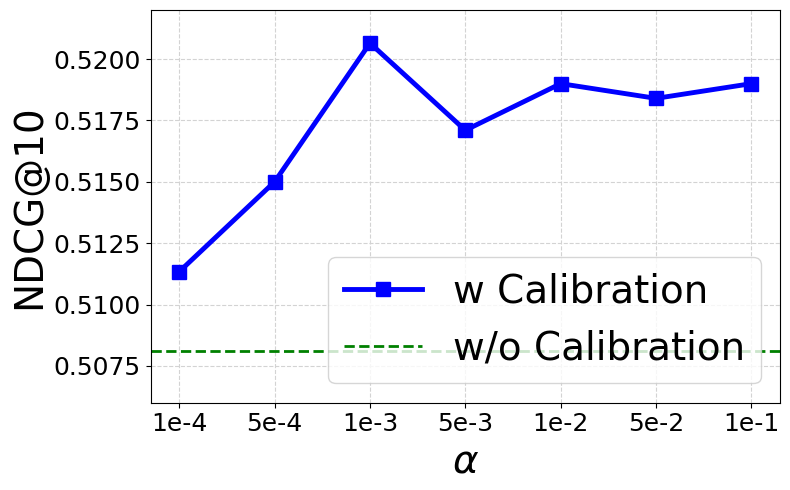}
    \subcaption{Impact of $\alpha$ on NDCG@10}
    \end{subfigure}
    \begin{subfigure}[b]{0.24\linewidth}
    \centering
      \includegraphics[width=\linewidth]{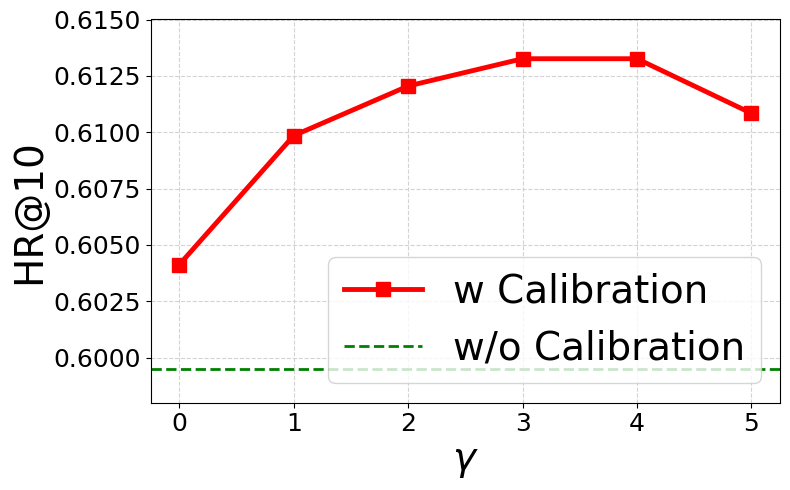}
      \subcaption{Impact of $\gamma$  on HR@10}
    \end{subfigure}
    \begin{subfigure}[b]{0.24\linewidth}
    \centering
    \includegraphics[width=\linewidth]{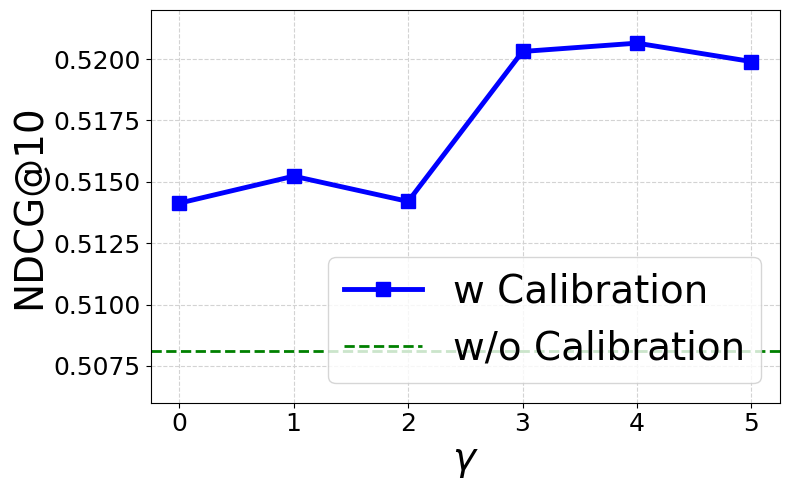}
    \subcaption{Impact of $\gamma$  on NDCG@10}
    \end{subfigure}
\caption{The hyper-parameter experiments on the weight $\alpha$ of $\mathcal{L}_\mathrm{cal}$, and the weight $\gamma$ of association score.  The results are based on the Fashion dataset with SASRec  model.
}
\label{fig:loss}

\end{figure*}

\subsection{Performance Comparison (RQ1)}
Table \ref{tab:Overall} presents the main experimental results. 
It demonstrates that \ours consistently outperforms recent state-of-the-art baselines across various backbone models,
confirming the generalizability and  effectiveness. 
A detailed analysis reveals that LLM-based methods consistently outperform traditional baselines, which primarily depend on interaction-based signals and attempt to address sparsity by augmenting rare items with popular ones.  
This observation underscores the significant benefit of incorporating semantic knowledge from pre-trained LLMs.
Among the traditional baselines, MELT often performs the best, whereas CITIES occasionally underperforms relative to backbone models, likely due to the seesaw effect, where gains on rare items are offset by considerable performance degradation on popular items.
Within the language-based methods, RLMRec frequently underperforms as it utilizes LLM as an auxiliary loss, failing to effectively leverage semantic representations. Although other recent LLM-based methods exhibit notable improvements over traditional approaches, they still fall short of \our. 
This highlights the distinctive advantage of \ours in   utilizing semantic information within LLM representations.

\subsection{Ablation Study (RQ2-a)}
To evaluate the individual effectiveness of each component, we conduct the ablation study and present the results in Table \ref{tab:exp_ablation}.
First, we analyze the impact by creating variants that remove the prefill embedding or the decoding embedding,  denoted as \textit{w/o Pref.} and 
\textit{w/o Decod.}, respectively.
The results demonstrate that removing either embedding degrades performance, which highlights the unique contribution of each component.
Moreover, the variant that eliminates the behavioral calibration objective, denoted as \textit{w/o Calib.}, also leads to reduced performance on all evaluated metrics, demonstrating the effectiveness of behavior awareness for tuning of the representation distribution.
The results of these three variants validate the design motivation of each component in \our.

\begin{table}[t]
\caption{Performance comparisons with different LLMs and sequence models on the Fashion dataset.}
        \vspace{-1pt}
\centering
\resizebox{0.87\linewidth}{!}{%
\begin{tabular}{ll|ccc}
\toprule
\multirow{2}{*}{\textbf{Backbone}} & \multirow{2}{*}{\textbf{Metric}} & \multicolumn{3}{c}{\textbf{LLM}} \\ 
\cmidrule{3-5}
                &           & LLama2 & LLama3.1 & Qwen2.5 \\ 
\midrule

\multirow{2}{*}{GRU4Rec}  & HR@10   & 0.5762  & 0.5698   & 0.5787 \\ 
                          & NDCG@10 & 0.4805  & 0.4783   & 0.4794 \\ 
\midrule
\multirow{2}{*}{Bert4Rec} & HR@10   & 0.5558  & 0.5500   & 0.5571 \\ 
                          & NDCG@10 & 0.4463  & 0.4524   & 0.4512 \\ 

\midrule
\multirow{2}{*}{SASRec}   & HR@10   & 0.6133  & 0.6151   & 0.6156 \\ 
                          & NDCG@10 & 0.5207  & 0.5190   & 0.5226 \\

\bottomrule
\end{tabular}%
}

\label{tab:LLMs_all_backbones}
\vspace{-5pt}
\end{table}

\subsection{Hyper-parameter Analysis (RQ2-b)}
To investigate the effects of behavioral calibration objective weighting $\alpha$ and behavioral signal weighting $\gamma$  in \our,
we conduct experiments on Fashion dataset with SASRec model, and present the overall performance trends in Figure ~\ref{fig:loss}. 
The hyper-parameter $\alpha$ determines the influence of behavioral calibration in the optimization process. As $\alpha$ increases, recommendation accuracy first improves and then declines. An excessively large $\alpha$ over-prioritizes behavior information, hindering the convergence, while an overly small $\alpha$ weakens its benefits, leading to suboptimal performance.
This underscores the importance of the proposed behavioral calibration objective. 
We further fix the weight of the behavioral calibration objective and adjust the value of the behavioral signal $\gamma$. The results, as shown in Figure~\ref{fig:loss}(c) and Figure~\ref{fig:loss}(d), also exhibit an increasing and then decreasing trend in the overall performance.  This reflects the subtle effect of either excessive or insignificant weights on the embedding distribution constraints. 
% Overall, once the weights are established in the appropriate range, \ours achieves significant and consistent performance improvements.
Overall, our approach achieves sustained performance improvements over a wide range of weights.

\subsection{Impact of Different LLMs (RQ3)}
To assess the adaptability of \our, we experiment with different base LLMs, including
LLama 2-7B-Chat~\cite{touvron2023llama}, LLama 3.1-8B-Instruct~\cite{grattafiori2024llama} and Qwen 2.5-7B-Instruct~\cite{qwen2.5}.
Results are presented in Table~\ref{tab:LLMs_all_backbones}.
Our analysis indicates that transitioning from LLama 2-7B-Chat to its successor LLama 3.1-8B-Instruct yields largely comparable performance, with only minor and inconsistent variations across metrics. 
And Qwen 2.5-7B-Instruct model delivers more  substantial gains, consistently outperforms in most configurations.
Overall, the stable  performance across different LLMs validates the  adaptability of our approach.

\begin{figure*}[ht]
\centering
    \begin{subfigure}[b]{0.24\linewidth}
    \centering
      \includegraphics[width=\linewidth, trim=60 0 60 0, clip]{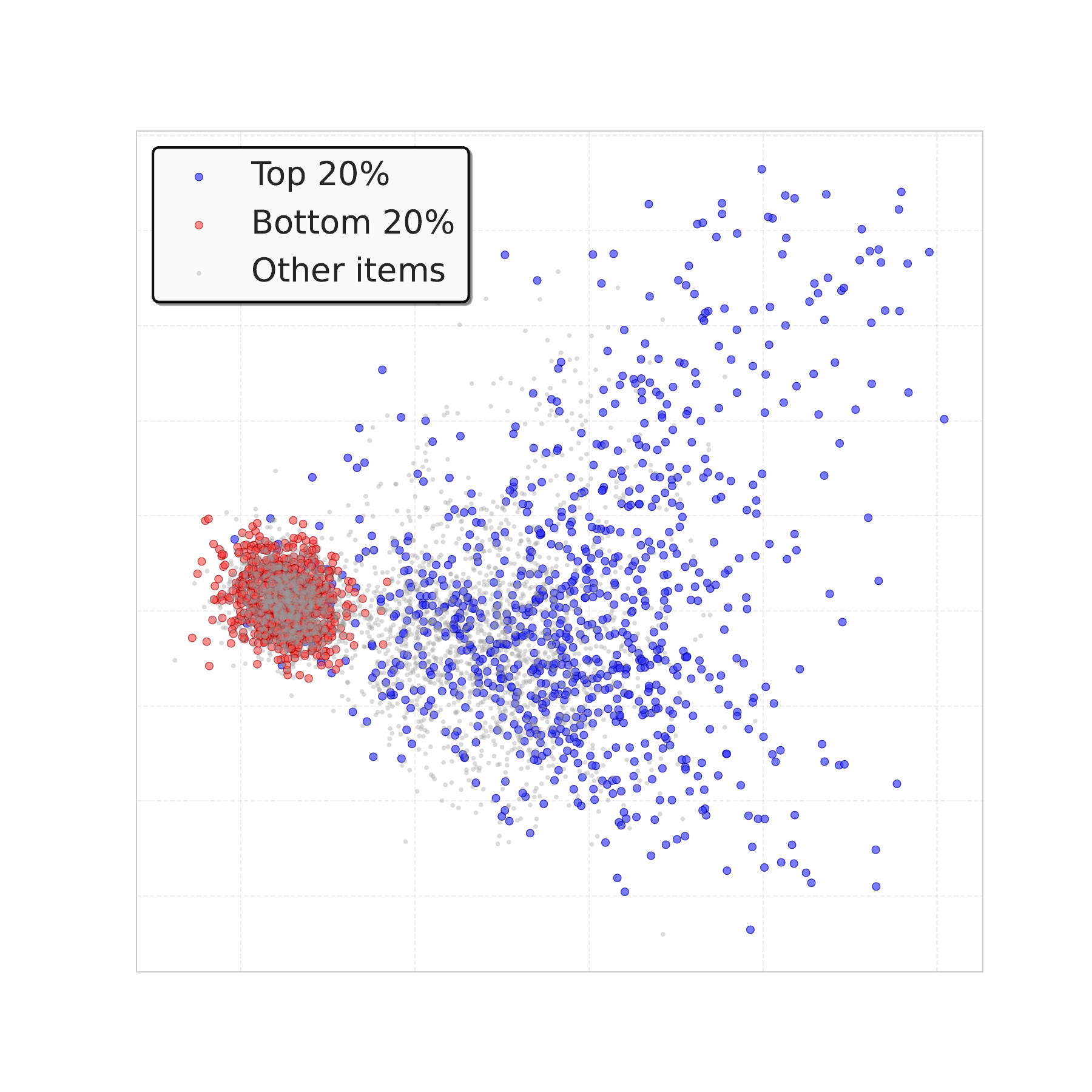}
      \subcaption{SASRec}

    \end{subfigure}
    \begin{subfigure}[b]{0.24\linewidth}
    \centering
      \includegraphics[width=\linewidth, trim=60 0 60 0, clip]{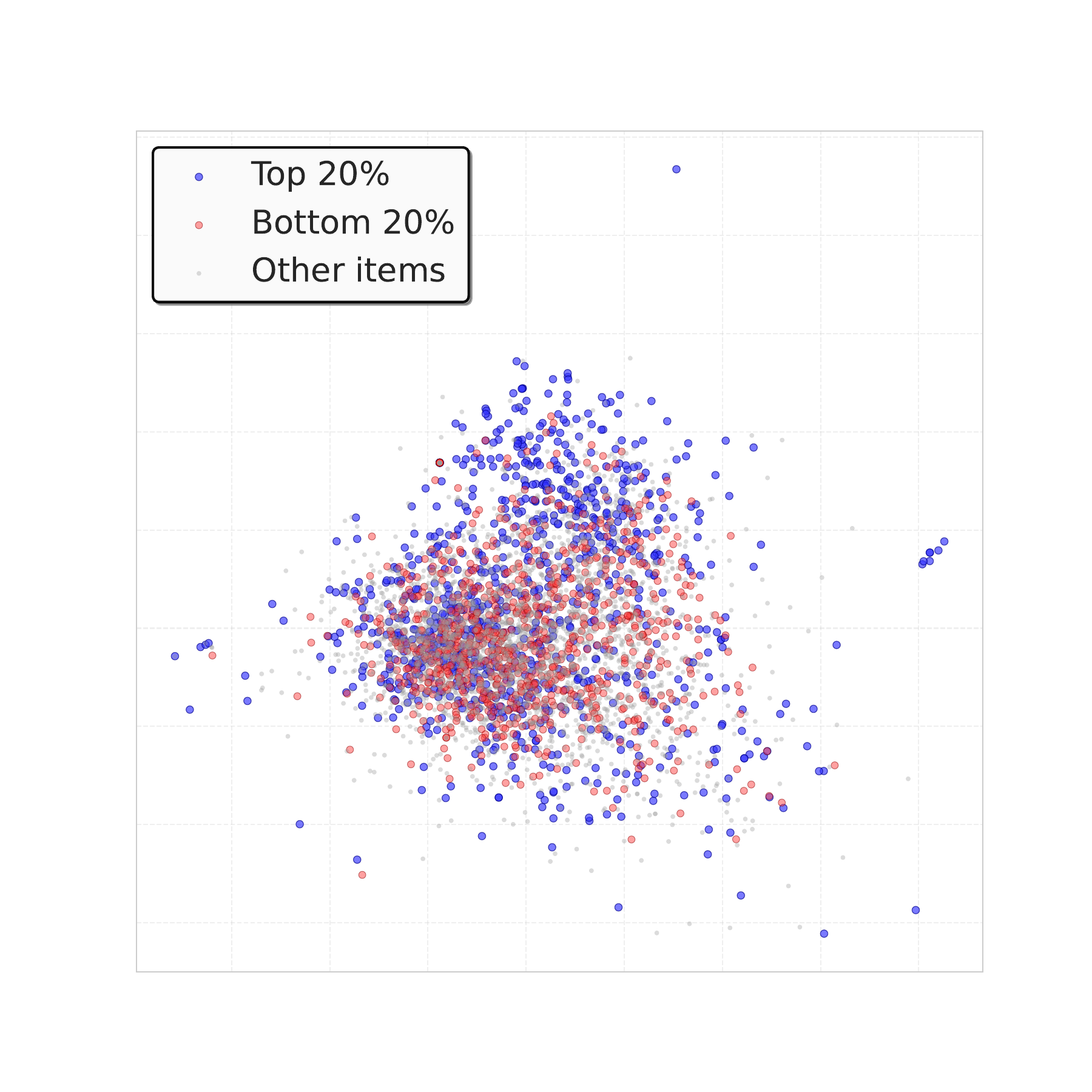}
    \subcaption{LLMEmb}

    \end{subfigure}
% \vspace{2mm}
    \begin{subfigure}[b]{0.24\linewidth}
    \centering
      \includegraphics[width=\linewidth, trim=60 0 60 0, clip]{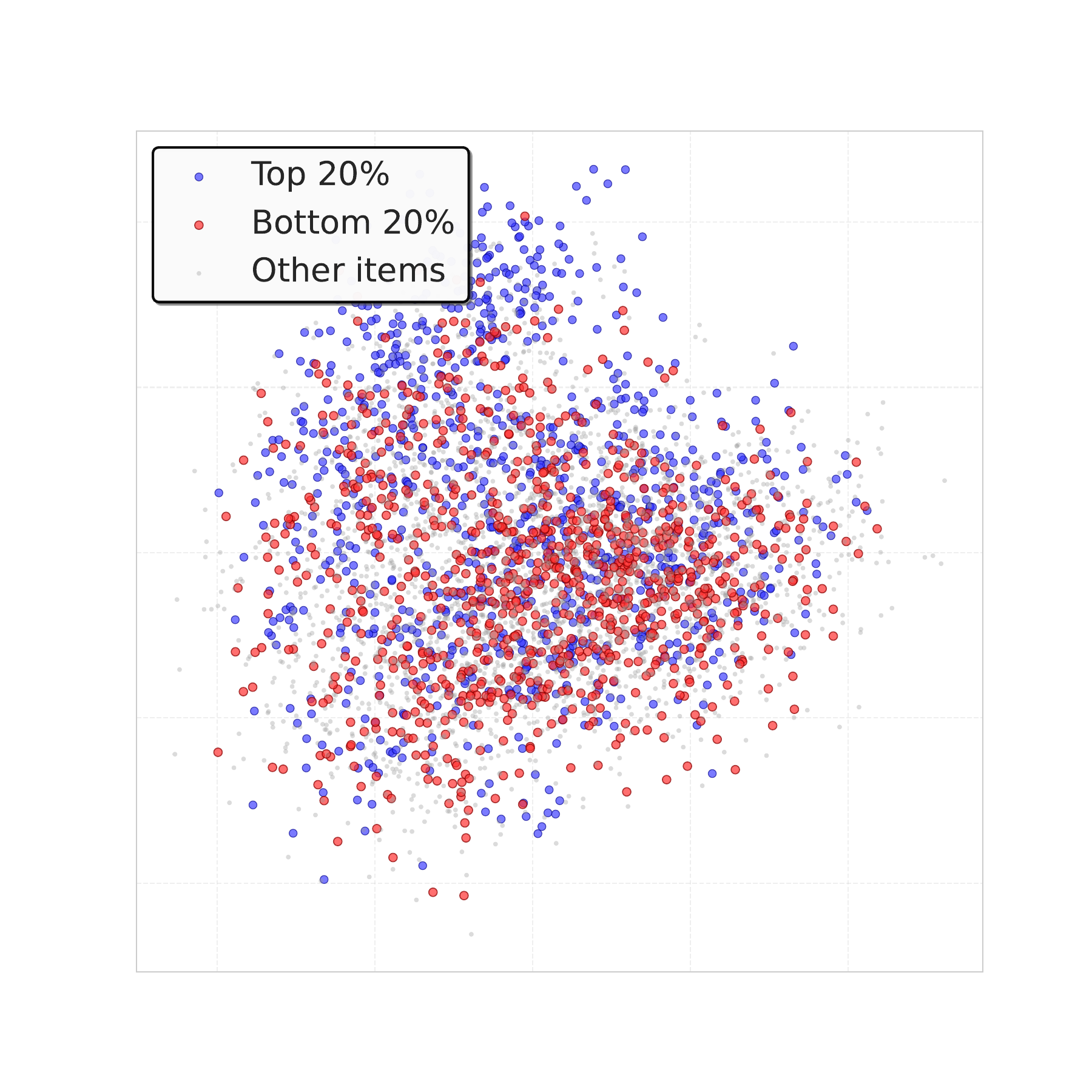}
      \subcaption{\our}

    \end{subfigure}
% \hfill %% useful if width of each figure is less the .5\textwidth
    \begin{subfigure}[b]{0.24\linewidth}
      \centering
      \raisebox{5mm}{
        \includegraphics[width=\linewidth, trim=0 0 0 0, clip]{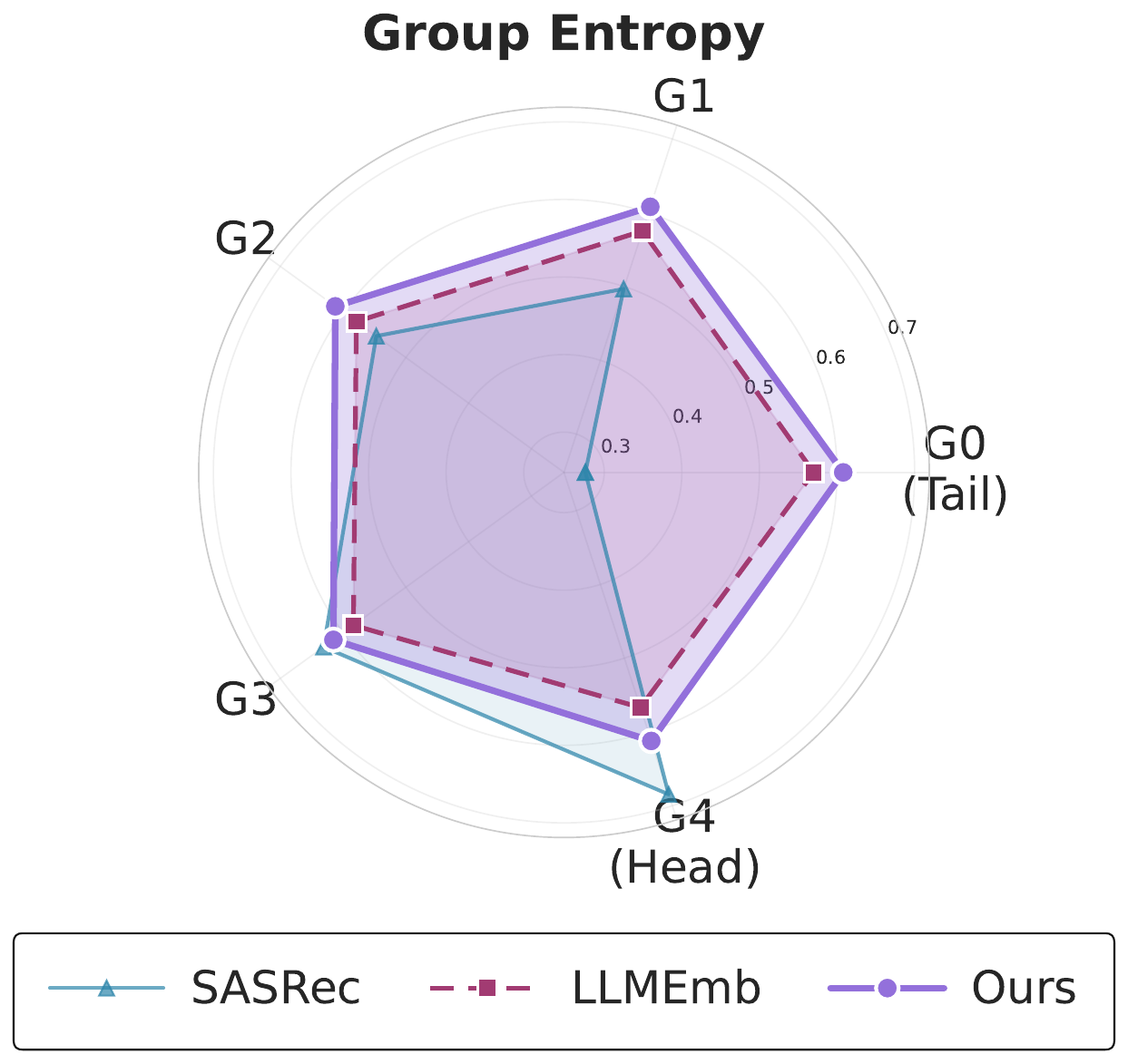}
      }
      \subcaption{Group Entropy}

    \end{subfigure}
\caption{The visualization of embeddings and group entropy.}
\label{fig:group}
\end{figure*}

\begin{table}[t]
    \caption{Training and Inference efficiency comparison.}
            \vspace{-1pt}
    % \vspace{-10pt}
    \centering

    \begin{tabular}{lcc}
        \toprule
        Method & Trainable Parameters & Inference GFLOPs\\
        \midrule
        SASRec & 5.72M &0.22 \\
        LLM-ESR &         6.98M &3.34 \\
        LLMEmb & 26.99M &0.22 \\
        AlphaFuse &5.72M &0.22 \\
        \ours & 5.95M &0.22\\
        \bottomrule
    \end{tabular}
    \label{tab:trainable_params}
                % \vspace{-5pt}
\end{table}

\subsection{Efficiency Analysis (RQ4)}
We compare the efficiency of \ours against SASRec and the latest baselines based on trainable parameters and inference GFLOPs. As shown in Table \ref{tab:trainable_params}, our proposed method achieves both training and inference efficiency. 
During training, compared to SASRec and AlphaFuse, \ours introduces only a small parameter overhead while requiring significantly fewer parameters than LLMEmb, which involves LLM fine-tuning. 
Moreover, during inference, \ours maintains the same  cost as efficient baselines, avoiding the additional computational burden introduced by LLM-ESR.
This demonstrates the superior  efficiency and deployment practicality of GenAIR.

\subsection{Group Analysis (RQ5)}
To further investigate how our proposed method affects different item groups, we divided items in the fashion dataset into five groups based on popularity. 
Figure~\ref{fig:group} shows the PCA two-dimensional visualization of the item embeddings for the most popular group (top 20\%) and the least popular group (bottom 20\%) with SASRec as the backbone model.
We observe significant differences in embedding distributions for long-tail items: SASRec’s embeddings are highly concentrated within a small region, while GenAIR’s are relatively dispersed.
We attribute this to the utilization of richer semantic and behavioral information, which aids in distinguishing diverse long-tail items and preventing  representation collapse.
To provide a more fundamental quantitative analysis, we further employ \textit{matrix-based entropy}~\cite{giraldo2014measures, skean2023dime} to measure the degree of information retention in each group, and present the average group entropy in Figure~\ref{fig:group}. Our findings demonstrate that GenAIR delivers more informative representations across most groups. Notably, in the most popular group, the values of LLMEmb and \ours are relatively lower than SASRec, an observation coherent with the see-saw effect. Detailed calculations for matrix entropy are provided in Appendix~\ref{sec:Entropy}.

\section{Related Work}

\label{Related Work}

\subsection{Sequential Recommendation}
Sequential recommendation models have gained wide attention for their ability to predict the next item a user will interact with~\cite{liu2023disentangling, liu2023diffusion,  liu2024sequential,  li2023automlp,  liu2023multi,  zhou2022filter, tang2018personalized, liu2021contrastive, kang2018self}.
Early approaches mainly focus on  neural  architecture design.
Caser ~\cite{tang2018personalized} uses convolutional neural networks  to model sequence patterns;
GRU4Rec ~\cite{hidasi2015session}  uses the gated recurrent unit;
 SASRec ~\cite{kang2018self} introduces a self-attention mechanism;
and Bert4Rec ~\cite{sun2019bert4rec} employs  a bidirectional encoder.
Subsequent studies have explored methods to improve  efficiency ~\cite{zhou2022filter}, and also strategies to mitigate the data sparsity issue ~\cite{liu2021contrastive, jang2020cities, kim2023melt}.
For instance,
CITIES ~\cite{jang2020cities} designs an embedding-inference function, and
MELT ~\cite{kim2023melt} employs a bilateral branching framework.
The latest advances have paid much attention to new paradigms and scale-up strategies ~\cite{yang2023generate, zhai2024actions, zhang2024wukong}.
DreamRec and DiffuASR ~\cite{yang2023generate, liu2023diffusion} reformulate the recommendation task through diffusion modeling;
HSTU~\cite{zhai2024actions} proposes a large-scale  generative model;
and Wukong~\cite{zhang2024wukong} explores the scaling law in recommendation.
While they have driven significant progress, a critical challenge persists: 
embedding layers are inherently bound to skewed or imbalanced distributions of interaction sequences,
which limits their capacity to capture subtle item representations and leads to suboptimal performance.

\subsection{LLM Embeddings for Recommendation}
Large language models~\cite{touvron2023llama, achiam2023gpt,qwen2.5} demonstrate great potential in enhancing recommendation~\cite{hou2024bridging, liu2025survey,luo2024integrating,luo2025recranker}.
To reduce computational cost and avoid using LLMs for frequent inference, a growing body of research has explored the use of LLM embeddings.
Generally, these efforts can be divided into two categories: 
(1) Text-based tuning: These works  ~\cite{harte2023leveraging, liu2024practice, liu2025llmemb, li2024calrec} enhance LLMs' understanding of recommendation scenarios through specifically designed fine-tuning frameworks. For instance,
TSLRec ~\cite{liu2024practice} designs a two-stage LLM training framework.
CALRec ~\cite{li2024calrec} introduces a two-stage fine-tuning framework using contrastive and language modeling losses. 
And LLMEmb ~\cite{liu2025llmemb} employs a fine-tuning strategy, using data augmentation and contrast learning to make the LLM more recommendation-friendly.
(2) Behavior-based tuning: These methods ~\cite{ren2024representation, hu2024enhancing,liu2024llmesr} design alignment mechanisms to match LLM-based semantic embeddings with ID-based embeddings.
SAID ~\cite{hu2024enhancing} feeds ID embeddings into the LLM and uses a projection module with a designed  text generation task.
RLMRec~\cite{ren2024representation} leverages LLM-derived semantic embeddings, mapping them to ID-based embeddings with an additional loss function.
 LLM-ESR ~\cite{liu2024llmesr} proposes a dual-view modeling framework that combines semantic  embeddings with ID-based embeddings via an attention mechanism. In addition, Alphafuse~\cite{DBLP:conf/sigir/Hu0LC0025} optimizes ID embeddings within the null space of semantic embeddings. 
 These methods, however,  fail to fully utilize LLM generative capabilities.

\section{Conclusion}
\label{conclusion}
In this work, we propose \our, a general framework for sequential recommendation with generative archetype-grounded item representations.
We leverage LLMs to generate archetypes from item metadata, thereby producing semantic embeddings that capture the target audience profiles.
Further, we introduce a behavioral calibration objective to ground these generative representations in real interaction patterns. 
This design effectively bridges the semantic richness of LLM knowledge with the behavioral reality of preferences.
The proposed framework is model-agnostic and  integrates seamlessly with sequential recommendation models. Validated by extensive experiments, GenAIR offers a principled approach for incorporating external knowledge into recommendation.

% \balance
\bibliographystyle{ACM-Reference-Format}
\balance
\bibliography{ourRef}

@article{achiam2023gpt,
  title={Gpt-4 technical report},
  author={Achiam, Josh and Adler, Steven and Agarwal, Sandhini and Ahmad, Lama and Akkaya, Ilge and Aleman, Florencia Leoni and Almeida, Diogo and Altenschmidt, Janko and Altman, Sam and Anadkat, Shyamal and others},
  journal={arXiv preprint arXiv:2303.08774},
  year={2023}
}

@inproceedings{behnamghaderllm2vec,
  title={LLM2Vec: Large language models are secretly powerful text encoders},
  author={BehnamGhader, Parishad and Adlakha, Vaibhav and Mosbach, Marius and Bahdanau, Dzmitry and Chapados, Nicolas and Reddy, Siva},
  booktitle={First Conference on Language Modeling},
year={2024}
}

@inproceedings{cao2022contrastive,
  title={Contrastive cross-domain sequential recommendation},
  author={Cao, Jiangxia and Cong, Xin and Sheng, Jiawei and Liu, Tingwen and Wang, Bin},
  booktitle={Proceedings of the 31st ACM International Conference on Information and Knowledge Management},
  pages={138--147},
  year={2022}
}

@article{fang2020deep,
 author = {Fang, Hui and Zhang, Danning and Shu, Yiheng and Guo, Guibing},
 journal = {ACM Transactions on Information Systems},
 pages = {1--42},
 title = {Deep learning for sequential recommendation: Algorithms, influential factors, and evaluations},
 year = {2020}
}

@article{guo2025deepseek,
  title={Deepseek-r1: Incentivizing reasoning capability in llms via reinforcement learning},
  author={Guo, Daya and Yang, Dejian and Zhang, Haowei and Song, Junxiao and Zhang, Ruoyu and Xu, Runxin and Zhu, Qihao and Ma, Shirong and Wang, Peiyi and Bi, Xiao and others},
  journal={arXiv preprint arXiv:2501.12948},
  year={2025}
}

@inproceedings{harte2023leveraging,
  author       = {Jesse Harte and
                  Wouter Zorgdrager and
                  Panos Louridas and
                  Asterios Katsifodimos and
                  Dietmar Jannach and
                  Marios Fragkoulis},

  title        = {Leveraging large language models for sequential recommendation},
  booktitle    = {Proceedings of the 17th {ACM} Conference on Recommender Systems},
  pages        = {1096--1102},
  year         = {2023},
  bibsource    = {dblp computer science bibliography, https://dblp.org}
}

@inproceedings{he2017neural,
  title={Neural collaborative filtering},
  author={He, Xiangnan and Liao, Lizi and Zhang, Hanwang and Nie, Liqiang and Hu, Xia and Chua, Tat-Seng},
  booktitle={Proceedings of the 26th International Conference on World Wide Web},
  pages={173--182},
  year={2017}
}

@inproceedings{hidasi2015session,
  author       = {Bal{\'{a}}zs Hidasi and
                  Alexandros Karatzoglou and
                  Linas Baltrunas and
                  Domonkos Tikk},
  title        = {Session-based recommendations with recurrent neural networks},
  booktitle    = {4th International Conference on Learning Representations},
  year         = {2016},
  bibsource    = {dblp computer science bibliography, https://dblp.org}
}

@article{hou2024bridging,
  title={Bridging language and items for retrieval and recommendation},
  author={Hou, Yupeng and Li, Jiacheng and He, Zhankui and Yan, An and Chen, Xiusi and McAuley, Julian},
  journal={arXiv preprint arXiv:2403.03952},
  year={2024}
}

@inproceedings{hu2024enhancing,
 author = {Hu, Jun and Xia, Wenwen and Zhang, Xiaolu and Fu, Chilin and Wu, Weichang and Huan, Zhaoxin and Li, Ang and Tang, Zuoli and Zhou, Jun},
 booktitle = {Companion Proceedings of the ACM on Web Conference 2024},
 pages = {103--111},
 title = {Enhancing sequential recommendation via llm-based semantic embedding learning},
 year = {2024}
}

@article{jaech2024openai,
  title={Openai o1 system card},
  author={Jaech, Aaron and Kalai, Adam and Lerer, Adam and Richardson, Adam and El-Kishky, Ahmed and Low, Aiden and Helyar, Alec and Madry, Aleksander and Beutel, Alex and Carney, Alex and others},
  journal={arXiv preprint arXiv:2412.16720},
  year={2024}
}

@inproceedings{jang2020cities,
  title={Cities: Contextual inference of tail-item embeddings for sequential recommendation},
  author={Jang, Seongwon and Lee, Hoyeop and Cho, Hyunsouk and Chung, Sehee},
  booktitle={2020 IEEE International Conference on Data Mining},
  pages={202--211},
  year={2020}
}

@inproceedings{kang2018self,
  title={Self-attentive sequential recommendation},
  author={Kang, Wang-Cheng and McAuley, Julian},
  booktitle={2018 IEEE International Conference on Data Mining},
  pages={197--206},
  year={2018}
}

@inproceedings{kim2023melt,
  title={Melt: Mutual enhancement of long-tailed user and item for sequential recommendation},
  author={Kim, Kibum and Hyun, Dongmin and Yun, Sukwon and Park, Chanyoung},
  booktitle={Proceedings of the 46th International ACM SIGIR conference on Research and Development in Information Retrieval},
  pages={68--77},
  year={2023}
}

@inproceedings{li2023automlp,
 author = {Muyang Li and
Zijian Zhang and
Xiangyu Zhao and
Wanyu Wang and
Minghao Zhao and
Runze Wu and
Ruocheng Guo},
 booktitle = {Proceedings of the ACM Web Conference 2023},
 pages = {1190--1198},
 title = {AutoMLP: Automated {MLP} for sequential recommendations},
 year = {2023}
}

@inproceedings{li2024calrec,
 author = {Li, Yaoyiran and Zhai, Xiang and Alzantot, Moustafa and Yu, Keyi and Vuli{\'c}, Ivan and Korhonen, Anna and Hammad, Mohamed},
 booktitle = {Proceedings of the 18th ACM Conference on Recommender Systems},
 pages = {422--432},
 title = {Calrec: Contrastive alignment of generative llms for sequential recommendation},
 year = {2024}
}

@article{liu2021contrastive,
 author = {Liu, Zhiwei and Chen, Yongjun and Li, Jia and Yu, Philip S and McAuley, Julian and Xiong, Caiming},
  journal={arXiv preprint arXiv:2108.06479},
 title = {Contrastive self-supervised sequential recommendation with robust augmentation},
 year = {2021}
}

@inproceedings{liu2023diffusion,
  title={Diffusion augmentation for sequential recommendation},
  author={Liu, Qidong and Yan, Fan and Zhao, Xiangyu and Du, Zhaocheng and Guo, Huifeng and Tang, Ruiming and Tian, Feng},
  booktitle={Proceedings of the 32nd ACM International Conference on Information and Knowledge Management},
  pages={1576--1586},
  year={2023}
}

@article{liu2023disentangling,
 author = {Liu, Qidong and Tian, Feng and Zheng, Qinghua and Wang, Qianying},
 journal = {Knowledge and Information Systems},
 pages = {2645--2664},
 title = {Disentangling interest and conformity for eliminating popularity bias in session-based recommendation},
 year = {2023}
}

@inproceedings{liu2023multi,
 author = {Ziru Liu and
Jiejie Tian and
Qingpeng Cai and
Xiangyu Zhao and
Jingtong Gao and
Shuchang Liu and
Dayou Chen and
Tonghao He and
Dong Zheng and
Peng Jiang and
Kun Gai},
 booktitle = {Proceedings of the ACM Web Conference 2023},
 pages = {1273--1282},
 title = {Multi-Task Recommendations with Reinforcement Learning},
 year = {2023}
}

@inproceedings{liu2024llmesr,
 author = {Qidong Liu and
Xian Wu and
Yejing Wang and
Zijian Zhang and
Feng Tian and
Yefeng Zheng and
Xiangyu Zhao},
  booktitle={The Thirty-eighth Annual Conference on Neural Information Processing Systems},
 title = {{LLM-ESR:} Large language models enhancement for long-tailed sequential recommendation},
  pages={26701--26727},
 year = {2024}
}

@article{liu2024practice,
 author = {Liu, Dugang and Xian, Shenxian and Lin, Xiaolin and Zhang, Xiaolian and Zhu, Hong and Fang, Yuan and Chen, Zhen and Ming, Zhong},
  journal={arXiv preprint arXiv:2406.00333},
 title = {A practice-friendly two-stage LLM-enhanced paradigm in sequential recommendation},
 year = {2024}
}

@inproceedings{liu2024sequential,
  title={Sequential recommendation for optimizing both immediate feedback and long-term retention},
  author={Liu, Ziru and Liu, Shuchang and Zhang, Zijian and Cai, Qingpeng and Zhao, Xiangyu and Zhao, Kesen and Hu, Lantao and Jiang, Peng and Gai, Kun},
  booktitle={Proceedings of the 47th International ACM SIGIR Conference on Research and Development in Information Retrieval},
  pages={1872--1882},
  year={2024}
}

@inproceedings{liu2025llmemb,
  title={LLMEmb: Large language model can be a good embedding generator for sequential recommendation},
  author={Liu, Qidong and Wu, Xian and Wang, Wanyu and Wang, Yejing and Zhu, Yuanshao and Zhao, Xiangyu and Tian, Feng and Zheng, Yefeng},
  booktitle={Proceedings of the AAAI Conference on Artificial Intelligence},
  number={11},
  pages={12183--12191},
  year={2025}
}

@article{liu2025survey,
  title={A survey of personalized large language models: Progress and future directions},
  author={Liu, Jiahong and Qiu, Zexuan and Li, Zhongyang and Dai, Quanyu and Yu, Wenhao and Zhu, Jieming and Hu, Minda and Yang, Menglin and Chua, Tat-Seng and King, Irwin},
  journal={arXiv preprint arXiv:2502.11528},
  year={2025}
}

@inproceedings{mcauley2015image,
  title={Image-based recommendations on styles and substitutes},
  author={McAuley, Julian and Targett, Christopher and Shi, Qinfeng and Van Den Hengel, Anton},
  booktitle={Proceedings of the 38th international ACM SIGIR Conference on Research and Development in Information Retrieval},
  pages={43--52},
  year={2015}
}

@inproceedings{mettes2019hyperspherical,
 author = {Pascal Mettes and
Elise van der Pol and
Cees Snoek},
  booktitle={Proceedings of the 33rd International Conference on Neural Information Processing Systems},
 pages = {1485--1495},
 title = {Hyperspherical prototype networks},
 year = {2019}
}

@inproceedings{muennighoff2024generative,
 author = {Muennighoff, Niklas and Hongjin, SU and Wang, Liang and Yang, Nan and Wei, Furu and Yu, Tao and Singh, Amanpreet and Kiela, Douwe},
 booktitle = {ICLR 2024 Workshop: How Far Are We From AGI},
 title = {Generative representational instruction tuning},
 year = {2024}
}

@article{pearson1901liii,
  title={LIII. On lines and planes of closest fit to systems of points in space},
  author={Pearson, Karl},
  journal={The London, Edinburgh, and Dublin philosophical magazine and journal of science},
  pages={559--572},
  year={1901},
}

@article{qwen2.5,
    title   = {Qwen2.5 technical report}, 
    author  = {An Yang and Baosong Yang and Beichen Zhang and Binyuan Hui and Bo Zheng and Bowen Yu and Chengyuan Li and Dayiheng Liu and Fei Huang and Haoran Wei and Huan Lin and Jian Yang and Jianhong Tu and Jianwei Zhang and Jianxin Yang and Jiaxi Yang and Jingren Zhou and Junyang Lin and Kai Dang and Keming Lu and Keqin Bao and Kexin Yang and Le Yu and Mei Li and Mingfeng Xue and Pei Zhang and Qin Zhu and Rui Men and Runji Lin and Tianhao Li and Tingyu Xia and Xingzhang Ren and Xuancheng Ren and Yang Fan and Yang Su and Yichang Zhang and Yu Wan and Yuqiong Liu and Zeyu Cui and Zhenru Zhang and Zihan Qiu},
    journal = {arXiv preprint arXiv:2412.15115},
    year    = {2024}
}

@inproceedings{ren2024representation,
 author = {Xubin Ren and
Wei Wei and
Lianghao Xia and
Lixin Su and
Suqi Cheng and
Junfeng Wang and
Dawei Yin and
Chao Huang},
 booktitle = {Proceedings of the ACM on Web Conference 2024},
 pages = {3464--3475},
 title = {Representation learning with large language models for recommendation},
 year = {2024}
}

@inproceedings{sun2019bert4rec,
  title={BERT4Rec: Sequential recommendation with bidirectional encoder representations from transformer},
  author={Sun, Fei and Liu, Jun and Wu, Jian and Pei, Changhua and Lin, Xiao and Ou, Wenwu and Jiang, Peng},
  booktitle={Proceedings of the 28th ACM International Conference on Information and Knowledge Management},
  pages={1441--1450},
  year={2019}
}

@inproceedings{tang2018personalized,
 author = {Jiaxi Tang and
Ke Wang},
  booktitle={Proceedings of the eleventh ACM International Conference on Web Search and Data Mining},
 pages = {565--573},
 title = {Personalized top-n sequential recommendation via convolutional sequence embedding},
 year = {2018}
}

@article{tang2024pooling,
 author = {Tang, Yixuan and Yang, Yi},
  journal={arXiv preprint arXiv:2409.02727},
 title = {Pooling and attention: What are effective designs for llm-based embedding models?},
 year = {2024}
}

@article{touvron2023llama,
 author = {Touvron, Hugo and Martin, Louis and Stone, Kevin and Albert, Peter and Almahairi, Amjad and Babaei, Yasmine and Bashlykov, Nikolay and Batra, Soumya and Bhargava, Prajjwal and Bhosale, Shruti and others},
  journal={arXiv preprint arXiv:2307.09288},
 title = {Llama 2: Open foundation and fine-tuned chat models},
 year = {2023}
}

@inproceedings{wang2020understanding,
 author = {Tongzhou Wang and
Phillip Isola},
  booktitle={Proceedings of the 37th International Conference on Machine Learning},
 pages = {9929--9939},
 title = {Understanding contrastive representation learning through alignment and uniformity on the hypersphere},
 year = {2020}
}

@inproceedings{xu2018spherical,
  title={Spherical latent spaces for stable variational autoencoders},
  author={Xu, Jiacheng and Durrett, Greg},
  booktitle={Proceedings of the 2018 Conference on Empirical Methods in Natural Language Processing},
  pages={4503--4513},
  year={2018}
}

@inproceedings{xu2024rethinking,
 author = {Wujiang Xu and
Qitian Wu and
Runzhong Wang and
Mingming Ha and
Qiongxu Ma and
Linxun Chen and
Bing Han and
Junchi Yan},
 booktitle = {Proceedings of the ACM on Web Conference 2024},
 pages = {3173--3184},
 title = {Rethinking cross-domain sequential recommendation under open-world assumptions},
 year = {2024}
}

@article{yang2023generate,
  title={Generate what you prefer: Reshaping sequential recommendation via guided diffusion},
  author={Yang, Zhengyi and Wu, Jiancan and Wang, Zhicai and Wang, Xiang and Yuan, Yancheng and He, Xiangnan},
  journal={Advances in Neural Information Processing Systems},
  volume={36},
  pages={24247--24261},
  year={2023}
}

@inproceedings{zhai2024actions,
  title={Actions speak louder than words: trillion-parameter sequential transducers for generative recommendations},
  author={Zhai, Jiaqi and Liao, Lucy and Liu, Xing and Wang, Yueming and Li, Rui and Cao, Xuan and Gao, Leon and Gong, Zhaojie and Gu, Fangda and He, Jiayuan and others},
  booktitle={Proceedings of the 41st International Conference on Machine Learning},
  pages={58484--58509},
  year={2024}
}

@inproceedings{zhang2024wukong,
  title={Wukong: towards a scaling law for large-scale recommendation},
  author={Zhang, Buyun and Luo, Liang and Chen, Yuxin and Nie, Jade and Liu, Xi and Li, Shen and Zhao, Yanli and Hao, Yuchen and Yao, Yantao and Wen, Ellie Dingqiao and others},
  booktitle={Proceedings of the 41st International Conference on Machine Learning},
  pages={59421--59434},
  year={2024}
}

@article{zhao2023embedding,
  title={Embedding in recommender systems: A survey},
  author={Zhao, Xiangyu and Wang, Maolin and Zhao, Xinjian and Li, Jiansheng and Zhou, Shucheng and Yin, Dawei and Li, Qing and Tang, Jiliang and Guo, Ruocheng},
  journal={arXiv preprint arXiv:2310.18608},
  year={2023}
}

@inproceedings{zhou2022filter,
  title={Filter-enhanced MLP is all you need for sequential recommendation},
  author={Zhou, Kun and Yu, Hui and Zhao, Wayne Xin and Wen, Ji-Rong},
  booktitle={Proceedings of the ACM Web Conference 2022},
  pages={2388--2399},
  year={2022}
}

@inproceedings{qu2024elephant,
  title={The elephant in the room: rethinking the usage of pre-trained language model in sequential recommendation},
  author={Qu, Zekai and Xie, Ruobing and Xiao, Chaojun and Kang, Zhanhui and Sun, Xingwu},
  booktitle={Proceedings of the 18th ACM Conference on Recommender Systems},
  pages={53--62},
  year={2024}
}

@inproceedings{DBLP:conf/sigir/Hu0LC0025,
  author       = {Guoqing Hu and
                  An Zhang and
                  Shuo Liu and
                  Zhibo Cai and
                  Xun Yang and
                  Xiang Wang},
  title        = {AlphaFuse: Learn {ID} embeddings for sequential recommendation in  null space of language embeddings},
  booktitle    = {Proceedings of the 48th International {ACM} {SIGIR} Conference on
                  Research and Development in Information Retrieval},
  pages        = {1614--1623},
  year         = {2025},
}

@article{giraldo2014measures,
  title={Measures of entropy from data using infinitely divisible kernels},
  author={Giraldo, Luis Gonzalo Sanchez and Rao, Murali and Principe, Jose C},
  journal={IEEE Transactions on Information Theory},
  pages={535--548},
  year={2014},
  publisher={IEEE}
}

@article{skean2023dime,
  title={{DiME}: Maximizing mutual information by a difference of matrix-based entropies},
  author={Skean, Oscar and Osorio, Jhoan Keider Hoyos and Brockmeier, Austin J and Giraldo, Luis Gonzalo Sanchez},
  journal={arXiv preprint arXiv:2301.08164},
  year={2023}
}

@article{kotler1969broadening,
  title={Broadening the concept of marketing},
  author={Kotler, Philip and Levy, Sidney J},
  journal={Journal of Marketing},
  pages={10--15},
  year={1969},
  publisher={SAGE Publications Sage CA: Los Angeles, CA}
}

@article{grattafiori2024llama,
  title={The llama 3 herd of models},
  author={Grattafiori, Aaron and Dubey, Abhimanyu and Jauhri, Abhinav and Pandey, Abhinav and Kadian, Abhishek and Al-Dahle, Ahmad and Letman, Aiesha and Mathur, Akhil and Schelten, Alan and Vaughan, Alex and others},
  journal={arXiv preprint arXiv:2407.21783},
  year={2024}
}

@inproceedings{chen2023star,
title={Topological representation learning for e-commerce shopping behaviors},
author={Chen, Yankai and Truong, Quoc-Tuan and Shen, Xin and Wang, Ming and Li, Jin and Chan, Jim and King, Irwin},
journal={Proceedings of the 19th International Workshop on Mining and Learning with Graphs},
year={2023}
}

@inproceedings{DBLP:conf/kdd/0001TS0K24,
  author       = {Yankai Chen and
                  Quoc{-}Tuan Truong and
                  Xin Shen and
                  Jin Li and
                  Irwin King},
  title        = {Shopping trajectory representation learning with pre-training for e-commerce customer understanding and recommendation},
  booktitle    = {Proceedings of the 30th {ACM} {SIGKDD} Conference on Knowledge Discovery
                  and Data Mining},
  pages        = {385--396},
  year         = {2024},
  bibsource    = {dblp computer science bibliography, https://dblp.org}
}

@inproceedings{DBLP:conf/cikm/QiuZ0CLDK24,
  author       = {Zexuan Qiu and
                  Jieming Zhu and
                  Yankai Chen and
                  Guohao Cai and
                  Weiwen Liu and
                  Zhenhua Dong and
                  Irwin King},
  title        = {{EASE:} Learning lightweight semantic feature adapters from large  language models for {CTR} prediction},
  booktitle    = {Proceedings of the 33rd {ACM} International Conference on Information
                  and Knowledge Management},
  pages        = {4819--4827},
  year         = {2024},
  bibsource    = {dblp computer science bibliography, https://dblp.org}
}

@inproceedings{DBLP:conf/aaai/Zhang00FK24,
  author       = {Xinni Zhang and
                  Yankai Chen and
                  Chenhao Ma and
                  Yixiang Fang and
                  Irwin King},
  title        = {Influential exemplar replay for incremental learning in recommender   systems},
  booktitle    = {Thirty-Eighth {AAAI} Conference on Artificial Intelligence},
  pages        = {9368--9376},
  year         = {2024},
  bibsource    = {dblp computer science bibliography, https://dblp.org}
}

@inproceedings{li2025g,
  title={G-refer: Graph retrieval-augmented large language model for explainable recommendation},
  author={Li, Yuhan and Zhang, Xinni and Luo, Linhao and Chang, Heng and Ren, Yuxiang and King, Irwin and Li, Jia},
  booktitle={Proceedings of the ACM on Web Conference 2025},
  pages={240--251},
  year={2025}
}

@article{zhang2023beyond,
  title={Beyond co-occurrence: Multi-modal session-based recommendation},
  author={Zhang, Xiaokun and Xu, Bo and Ma, Fenglong and Li, Chenliang and Yang, Liang and Lin, Hongfei},
  journal={IEEE Transactions on Knowledge and Data Engineering},
  pages={1450--1462},
  year={2023},
  publisher={IEEE}
}

@inproceedings{wu2025ip2,
  title={IP2: Entity-Guided Interest Probing for Personalized News Recommendation},
  author={Wu, Youlin and Sun, Yuanyuan and Zhang, Xiaokun and Zhan, Haoxi and Xu, Bo and Yang, Liang and Lin, Hongfei},
  booktitle={Proceedings of the Nineteenth ACM Conference on Recommender Systems},
  pages={187--196},
  year={2025}
}

@article{zhang2025survey,
  title={A survey on side information-driven session-based recommendation: from a data-centric perspective},
  author={Zhang, Xiaokun and Xu, Bo and Li, Chenliang and He, Bowei and Lin, Hongfei and Ma, Chen and Ma, Fenglong},
  journal={IEEE Transactions on Knowledge and Data Engineering},
  year={2025},
  pages={1--20},
  publisher={IEEE}
}

@article{luo2024integrating,
  title={Integrating large language models into recommendation via mutual augmentation and adaptive aggregation},
  author={Luo, Sichun and Yao, Yuxuan and He, Bowei and Huang, Yinya and Zhou, Aojun and Zhang, Xinyi and Xiao, Yuanzhang and Zhan, Mingjie and Song, Linqi},
  journal={arXiv preprint arXiv:2401.13870},
  year={2024}
}

@article{luo2025recranker,
  title={Recranker: Instruction tuning large language model as ranker for top-k recommendation},
  author={Luo, Sichun and He, Bowei and Zhao, Haohan and Shao, Wei and Qi, Yanlin and Huang, Yinya and Zhou, Aojun and Yao, Yuxuan and Li, Zongpeng and Xiao, Yuanzhang and others},
  journal={ACM Transactions on Information Systems},
  pages={1--31},
  year={2025},
}

\appendix
\section{Case Study}
\label{appendix:example}

To illustrate the differences in archetype generation across different large language models, we present a case study analyzing outputs from \textbf{Llama 2-7B-Chat} and \textbf{Qwen 2.5-7B-Instruct}. Both models were tasked with generating user archetypes for a specific item: the \textit{Zutano Baby UPF 30+ Sun Hat}. This item originates from the Fashion dataset, with a clearly defined core context: it is an accessory designed specifically for infants and toddlers, emphasizing safety (UV protection) and aesthetics (``cute chin strap''). Analysis of its generated outputs (see Table \ref{user_archetype_item_25} for details) reveals significant differences in its contextual understanding and generative capabilities.

(\romannumeral 1) Llama 2-7B-Chat.
While Llama2's output covers ten distinct archetypes, it deviates significantly from the item's core context: \textbf{infants}.
First, archetype generation is inaccurate. For instance, several user types generated by the model are logically incompatible with the item. Examples include suggesting the hat would appeal to ``athletes, fitness enthusiasts, or those who engage in activities like running, playing sports, or doing yoga''.
Second, content hallucination. For instance, the model invents item attributes by stating ``...the breathable cotton material of this hat may offer relief...'' despite the source description offering no information about hat materials. The model fabricates plausible yet unverified details based on common knowledge about infant products, undermining modeling of the item's uniqueness.
Third, the content is overly broad. While some generated archetypes (e.g., ``parents or caregivers'') are accurate, they are diluted by predominantly irrelevant or flawed suggestions like ``outdoor enthusiasts'' or ``fashion-conscious individuals'', failing to explicitly state that these individuals must be parents, guardians, or purchasing for children.

(\romannumeral 2) Qwen 2.5-7B-Instruct.
Its responses are more concise, comprising only five archetypes, yet demonstrate superior contextual awareness and logical reasoning. The outputs are more relevant, reflecting a deeper understanding of the item and its target user groups.
First, contextual focus. Each archetype generated by Qwen 2.5 connects directly and logically to the core theme of early childhood item. It accurately identifies the primary users (``parents and guardians'') and meticulously expands relevant groups, such as ``grandparents'' (a significant gift-giving demographic) and ``Style-Conscious Parents".
Second, logical reasoning. The model effectively connects item features to user motivations. It associates the  ``UPF 30+ sun protection'' feature with ``Health-Conscious Individuals'' and correctly defines ``Outdoor Enthusiasts'' as ``Families who frequently engage in outdoor activities''. This demonstrates the model's ability to integrate multiple pieces of information into a coherent and accurate user profile.
Third, high-quality output. The generated archetype list is focused and accurate, with no noticeable noise or hallucinations, indicating a more robust and reliable reasoning process.

This comparative analysis clearly demonstrates the significant differences exhibited by LLMs with varying generative capabilities in reasoning and potential user group generation. The findings from this case also align with its application performance embedded in sequential recommendation.

\section{More Experiment Details}

\subsection{Proof of Proposition~\ref{prop:force-law}}
\label{appendix:proof1}

For a mini-batch of ordered pairs, we define the calibration loss as:
\begin{align}
\widehat{\mathcal L}_{\mathrm{cal}}
&=\frac1{|\mathcal B|}
\sum_{(i,j)\in\mathcal B}
w(i,j)\,
\exp \bigl(-t\lVert\mathbf e_i-\mathbf e_j\rVert_2^{2}\bigr), \label{eq:cal1} \\
&=\frac1{|\mathcal B|}
\sum_{(i,j)\in\mathcal B}
w(i,j)\,
\exp \bigl(\beta\,\mathbf e_i^{\!\top}\mathbf e_j\bigr), \label{eq:cal2}
\end{align}
where $\beta = 2t$ and the constant $e^{-2t}$ is omitted as it vanishes upon differentiation.
Fixing an index $i$, the vector $\mathbf{e}_i$ appears only in the summands where it is the first component of the ordered pair. Therefore, the gradient with respect to $\mathbf e_i$ is given by:

\begin{equation}
\label{eq:cal3}
\frac{\partial\widehat{\mathcal L}_{\mathrm{cal}}}{\partial\mathbf e_i}
=\frac1{|\mathcal B|}
\sum_{j:\,(i,j)\in\mathcal B}
w(i,j)\;
\frac{\partial}{\partial\mathbf e_i}
\exp \bigl(\beta\,\mathbf e_i^{\!\top}\mathbf e_j\bigr). \end{equation}
Applying the chain rule, we compute:

\begin{equation}
\frac{\partial}{\partial\mathbf e_i}
\exp \bigl(\beta\,\mathbf e_i^{\!\top}\mathbf e_j\bigr)
=
\beta\,\exp \bigl(\beta\,\mathbf e_i^{\!\top}\mathbf e_j\bigr)\,\mathbf e_j .
\end{equation}
Substituting this result back into Equation~\ref{eq:cal3}, we obtain:

\begin{equation}
\frac{\partial\widehat{\mathcal L}_{\mathrm{cal}}}{\partial\mathbf e_i}
=\frac{\beta}{|\mathcal B|}
\sum_{j:\,(i,j)\in\mathcal B}
w(i,j)\,
e^{\beta\,\mathbf e_i^{\!\top}\mathbf e_j}\,
\mathbf e_j,
\end{equation}
which directly yields the stated result.

\subsection{Details of Datasets}
\label{appendix:datasets}

\textbf{Yelp} contains the user's check-in history and corresponding reviews, where we use the attribute information and check-in data.
The Amazon datasets ~\cite{mcauley2015image} are e-commerce datasets obtained from Amazon platform with user ratings and reviews, and we adopt two of them to evaluate, namely \textbf{Beauty} and \textbf{Fashion}. 
The statistics of the datasets are summarized in Table~\ref{tab:statistics}.

\begin{table}[h]
\centering
\caption{The statistics of datasets after preprocessing.}

\tabcolsep=0.3cm   
\begin{tabular}{ccccc}
\toprule
\textbf{Dataset} & \textbf{Users} & \textbf{Items} & \textbf{Avg. Len} & \textbf{Sparsity}  \\ 
\midrule
Yelp & 15,720 & 11,383 & 12.23 & 99.89\%  \\
Beauty & 52,204 & 57,289 & 7.57 & 99.99\%  \\ 
Fashion & 9,049 & 4,722 & 3.82 & 99.92\%  \\
\bottomrule
\end{tabular}
\label{tab:statistics}

\end{table}

\subsection{Details of Backbones and Baselines}
\label{appendix:baselines}

\subsubsection{Backbones}

(\romannumeral 1) \textbf{GRU4Rec}~\cite{hidasi2015session} employs Gated Recurrent Units (GRUs) as the sequence encoder and utilizes a sequence-to-one pairwise ranking loss for optimizing the recommendation task.
(\romannumeral 2) \textbf{Bert4Rec}~\cite{sun2019bert4rec} introduces a hybrid training objective that combines pairwise ranking loss with cloze task, which involves masking a subset of items within a sequence, enabling the model to learn bidirectional contextual representations. 
(\romannumeral 3) \textbf{SASRec}~\cite{kang2018self} utilizes causal self-attention layers as the basic unit of its sequence encoder, and it is optimized using a sequence-to-sequence pairwise ranking loss, which aligns with its autoregressive architecture.

\subsubsection{Baselines.}
\noindent\textbf{Traditional methods.} This type of approach uses inherent behavioral information to achieve self-enhancement, typically enhancing long-tailed groups with representations of popular groups.
(\romannumeral 1) \textbf{CITIES}~\cite{jang2020cities} introduces an embedding-inference function trained on head items, which is then applied to refine long-tail item embeddings during the inference process.
(\romannumeral 2) \textbf{MELT}~\cite{kim2023melt} employs a bilateral branching architecture that separately processes long-tail and head users/items, effectively addressing the cold-start problem through targeted augmentation strategies.

\noindent\textbf{Language-based Method.} These methods augment recommendation models with semantic representations derived from LLMs.
(\romannumeral 1) \textbf{RLMRec}~\cite{ren2024representation} is an representative approach leveraging LLM-derived semantic embeddings for collaborative filtering. We refer to past implementations ~\cite{liu2024llmesr} to map semantic LLM embeddings to behavior space via a two-layer MLP, and use similarity differences as the designed additional loss function.
(\romannumeral 2)  \textbf{LLMInit}~\cite{harte2023leveraging, hu2024enhancing, qu2024elephant} uses semantic embeddings from LLM to initialize the item representations, a projector is applied to transforms the dimensions, ensuring compatibility with the recommendation backbone model and replacing the original ID-based embeddings.
(\romannumeral 3)  \textbf{LLM-ESR}~\cite{liu2024llmesr} proposes a dual-view modeling framework that combines semantic context-based embedding and ID-based embedding, enhances each other through an attention mechanism, and mitigates the long-tail problem through a retrieval-augmented self-distillation algorithm.
(\romannumeral 4)  \textbf{LLMEmb}~\cite{liu2025llmemb} uses a fine-tuning approach based on feature-level data augmentation and contrast learning to make the LLM more recommendation-friendly, and enhances the semantic embeddings using ID-based embeddings that are trained in advance.
(\romannumeral 5)  \textbf{Alphafuse}~\cite{DBLP:conf/sigir/Hu0LC0025} optimizes item representations in sequential recommendation by learning ID embeddings within the null space of language embeddings.

\subsection{Implementation Details}
\label{appendix:implementation}
The implementation is based on Python 3.9.5 and PyTorch 2.6.0. 
And the hardware used is an Intel(R) Xeon(R) Silver 4114 platform with NVIDIA RTX A6000 48GB GPUs.
For the backbones that have distinct architectures: GRU4Rec utilizes a single GRU layer, whereas SASRec and Bert4Rec implement two self-attention layers each. 
We fix the final embedding dimensionality to 128 for all methods. The dimension of the item embedding after the PCA step is 384.
For consistent training across all models, we maintain uniform hyper-parameters, with a batch size of 128 and a learning rate of 0.001. We evaluate the model performance using the NDCG@10 metric on the validation set for optimization, and adopt the early stopping strategy to avoid overfitting.

\subsection{Matrix-based Entropy}
\label{sec:Entropy}
To formally quantify the information content and richness of item representations, we utilize the concept of \emph{matrix-based entropy}~\cite{giraldo2014measures, skean2023dime}.
Let $\mathbf{Z} \in \mathbb{R}^{N \times D}$ denote a matrix containing $N$ item embeddings, where each item is represented by a $D$-dimensional vector. We first define the \emph{Gram matrix} as:
\begin{equation}
\mathbf{K} = \mathbf{Z}\mathbf{Z}^\top.
\end{equation}
Then the matrix-based $\rho$-order entropy is defined directly on the Gram matrix $\mathbf{K}$. For any order $\rho > 0$, it is given by:
\begin{equation}
\label{eq:Matrix-Based Entropy}
S_\rho(\mathbf{Z}) := \frac{1}{1-\rho} \log \biggl( \sum_{i=1}^{r} \Bigl(\tfrac{\lambda_i(\mathbf{K})}{\mathrm{tr}(\mathbf{K})}\Bigr)^\rho \biggr),
\end{equation}
where $r = \mathrm{rank}(\mathbf{K}) \leq \min(N,D)$ is the rank of the Gram matrix, and $\mathrm{tr}(\mathbf{K}) = \sum_i \lambda_i(\mathbf{K})$ is its trace.
Here, the order of $\rho$ allows for a smooth transition between different entropy definitions. 
For simplicity, we consistently use $\rho=1$ throughout this paper.

\begin{table*}[ht]
\caption{An Example of Latent User Archetype Generation}
\centering
\begin{tabularx}{\textwidth}{X}
\toprule

\textbf{\#\# Item \#\#} \\
Item ID: 25 \\
Item: B0009A1EA6 \\
\midrule

\textbf{\#\# Context \#\#} \\
Based on the following attributes, identify the types of users this item would appeal to: 

The item name is Zutano Baby UPF 30+ Sun Protection Hat.

The item description is "Our wide-brim sun hat with adorable chin ties will not only stay on but will keep your little one". 

The item feature is Select Size to see the return policy for the item; rank is 464,412 in Clothing, Shoes Jewelry.

The item average user rating is 5 star. \\
\midrule

\textbf{\#\# Generated Archetype Text (LLama 2-7B-Chat) \#\#} \\

Based on the provided information, the Zutano Baby UPF 30+ Sun Protection Hat is likely to appeal to specific user groups who prioritize safety, functionality, and style for young children.

1. Parents or caregivers of young children (ages 0–5): The product's focus on providing protection from the sun and its cute design make it an ideal choice for parents who want to keep their little ones safe while also dressing them in style.

2. Outdoor enthusiasts: People who enjoy spending time outdoors, such as hikers, bicyclists, or beachgoers, may appreciate the hat's ability to provide reliable sun protection without sacrificing style.

3. Fashion-conscious individuals: Although the hat is primarily marketed towards parents and caregivers, anyone looking for stylish headwear that also offers excellent sun protection could find this product appealing.

4. Individuals living in sunny climates: Residents of areas with high levels of UV radiation may prioritize sun protection when selecting clothing and accessories, making the Zutano Baby UPF 30+ Sun Protection Hat an attractive option.

5. Active people: This group includes athletes, fitness enthusiasts, or those who engage in activities like running, playing sports, or doing yoga, where they need to protect themselves from the sun during exercise.

6. Travelers: Whether visiting tropical destinations or enjoying summer vacation at home, travelers might seek out a practical yet stylish sun hat for their adventures.

7. Those with sensitive skin: Some individuals may have skin sensitivity or allergies that make wearing certain fabrics uncomfortable or irritating. In such cases, the breathable cotton material of this hat may offer relief and peace of mind.

8. Gift buyers: With its affordability, versatility, and cute design, this hat could become a popular gift item for friends, family members, or colleagues with young children.

9. Educators or daycare providers: Teachers, nannies, or other professionals responsible for supervising young children throughout the day may appreciate the convenience and effectiveness of this sun hat.

10. Photographers or influencers: Professionals working in creative fields where visual appearance matters may admire the Zutano Baby. \\

\midrule

\textbf{\#\# Generated Archetype Text (Qwen 2.5-7B-Instruct) \#\#} \\

The Zutano Baby UPF 30+ Sun Protection Hat appears to be a fashionable and practical accessory designed specifically for infants and young children. Based on its attributes, the following types of users would likely find this item appealing:

1. **Parents and Guardians**: This is the primary target audience as the hat is designed for babies and young children. Parents who are concerned about sun protection for their children would find this hat beneficial.

2. **Grandparents**: Grandparents often play a significant role in caring for grandchildren and may appreciate the sun protection features of this hat.

3. **Outdoor Enthusiasts**: Families who frequently engage in outdoor activities such as picnics, beach trips, or park visits would value a hat that keeps their children shaded and protected from the sun.

4. **Health-Conscious Individuals**: People who prioritize health and well-being might choose this hat to ensure their children receive adequate sun protection during outdoor activities.

5. **Style-Conscious Parents**: The wide-brim design and cute chin ties suggest that the hat is also stylish, which could appeal to parents who want both functionality and fashion for their children's accessories.

Given the high star rating (5 stars) and the specific mention of UPF 30+ sun protection, the hat is clearly designed with safety and comfort in mind, making it particularly suitable for families seeking reliable sun protection for their young ones. \\

\bottomrule

\end{tabularx}
\label{user_archetype_item_25}
\end{table*}

% \onecolumn

\end{document}